\documentclass{report}
 
\usepackage{amsmath}
\usepackage{algorithmic}
\usepackage{graphicx}
\usepackage{textcomp}
\usepackage{xcolor}
\usepackage{hyperref}
\usepackage{multirow}
\usepackage{subfigure}
\usepackage{enumitem}
\usepackage{color,soul}
\usepackage[flushleft]{threeparttable}
\usepackage[T1]{fontenc}
\usepackage{cite}
\usepackage{lscape} 
 \usepackage{lipsum}

\newcommand\blfootnote[1]{%
  \begingroup
  \renewcommand\thefootnote{}\footnote{#1}%
  \addtocounter{footnote}{-1}%
  \endgroup
}

\begin{document}
%
\title{  
  BGP-Multipath Routing in the  Internet
  \blfootnote{This work was supported by China Scholarship Council (CSC) with grant no. 201406060022.
{Jie Li and Shi Zhou are with  Department of Computer Science, {University College London} (UCL), UK (emails: jie.li@cs.ucl.ac.uk \& s.zhou@ucl.ac.uk).}
{Vasileios Giotsas is with School of Computing and Communications, Lancaster University, UK (email: v.giotsas@lancaster.ac.uk).}
{Yangyang Wang is with Institute for Network Sciences and Cyberspace, Beijing National Research Center for Information Science and Technology (BNRist), Tsinghua University, Beijing, China (email: wangyy@cernet.edu.cn).}
}
\blfootnote{Published in IEEE  Transactions on Network and Service Management (TNSM) in May 2022.}
\blfootnote{DOI: 10.1109/TNSM.2022.3177471}\\
%
}
%
%
%

\author{Jie Li, 
       Vasileios Giotsas, 
       Yangyang Wang %
        and Shi Zhou
}


\date{May 2022}




\maketitle


\begin{abstract}

BGP-Multipath (BGP-M) is a multipath routing technique for load balancing. 
Distinct  from other techniques  deployed at a router inside an Autonomous System (AS), BGP-M is deployed  at a border router 
that has installed multiple  inter-domain  border links  to a neighbour AS. 
It uses the equal-cost multi-path (ECMP) function of a border router  to share traffic to a destination prefix on different border links.
Despite recent research interests in multipath routing, there is little study on BGP-M.

Here we provide the first measurement and a comprehensive analysis of BGP-M routing in the Internet.   
We extracted information on BGP-M from query data collected from   Looking Glass (LG) servers.
We revealed that BGP-M has already been  extensively deployed and used in the Internet.
A particular example is Hurricane Electric (AS6939), a Tier-1 network operator, which has implemented >1,000 cases of BGP-M at 69   of its border routers to prefixes in  611  of its neighbour ASes, including many hyper-giant ASes and large content providers, on both IPv4 and IPv6 Internet.  
We examined the distribution and operation of BGP-M. We also   ran traceroute   using  RIPE Atlas  to infer  the routing paths, the schemes of traffic allocation, and  the delay on border links. 
This study provided  the state-of-the-art knowledge on BGP-M   with novel insights into the unique features  
and  the distinct advantages of BGP-M as an effective and readily available technique for  load balancing. 

 
\end{abstract}


%

\section{Introduction}
\label{Sec:Introduction}

The  default setting of Border  Gateway Protocol (BGP)~\cite{RFC4271} requires a single ``best'' path for each prefix. BGP-Multipath (BGP-M) is a technique to  enable load balancing on multiple IP-level inter-domain paths of equal cost.
Specifically, a network operator can   activate the Equal-Cost Multi-Path (ECMP) function at a border router so that when the border router learns from a  {same neighbour Autonomous System (AS)} multiple eBGP paths (via different border links) to a prefix with  equal attributes, the border router    installs all of these paths in the routing table instead of trying additional tie-breaking attributes. 
Routers produced by most major   vendors support the ECMP function, including  Juniper~\cite{juniper-mbgp},  Cisco~\cite{CISCO},  and  Huawei~\cite{huawei-mbgp}.
Although there have been a number of research works on multipath routing, e.g.~\cite{Augustin2011TON,  Almeida2017PAM, Vermeulen2018IMC, Vermeulen2020NSDI, Almeida2020INFOCOM},  
BGP-M remains an obscure technique.

In this paper, we present the first measurement and a comprehensive analysis on the  BGP-M routing in the Internet. We obtained BGP data from Looking Glass (LG) servers to infer the deployment of  BGP-M, and collected traceroute data from RIPE Atlas \cite{RIPE2015IPJ} to extract further details on BGP-M  routing paths, the schemes of traffic allocation, and  the delay on border links. 
Our results showed that BGP-M has been deployed extensively in the Internet.

The techniques and results presented in this paper provide the state-of-the-art knowledge on BGP-M.
We believe that our work is relevant to industry stakeholders, Internet engineers and researchers interested in Internet routing performance and security.

\begin{table}[b]
 
    \centering
        \caption{Notations and Descriptions }
    \label{tab:Notations}
    \begin{tabular}{l|l}
    \hline
     Notation       & Description\\ \hline
    {\em{SrcIP}}        & Source IP address \\
    {\em{DstIP}}        & Destination IP address \\
    {\em{DstPrfx}}      & Destination prefix   \\
     \hline
    {\em{NearAS}}       & Nearside AS \\
    {\em{NearBR}}      & Nearside border router \\
    {\em{NearIP}}       & IP address of ingress interface of {\em{NearBR}} \\
     \hline
    {\em{FarAS}}        & Farside AS \\
    {\em{FarBR}}       & Farside border router \\
    {\em{FarIP}}        & IP address of ingress interface of {\em{FarBR}}\\
     \hline
    {\em{BL}}           & Border link between ASes \\
     \hline
 
    \end{tabular}
 
\end{table}

\begin{figure}[t]
    \centering
    
    \subfigure[Best-path routing]{
    \includegraphics[width = 0.8\textwidth]{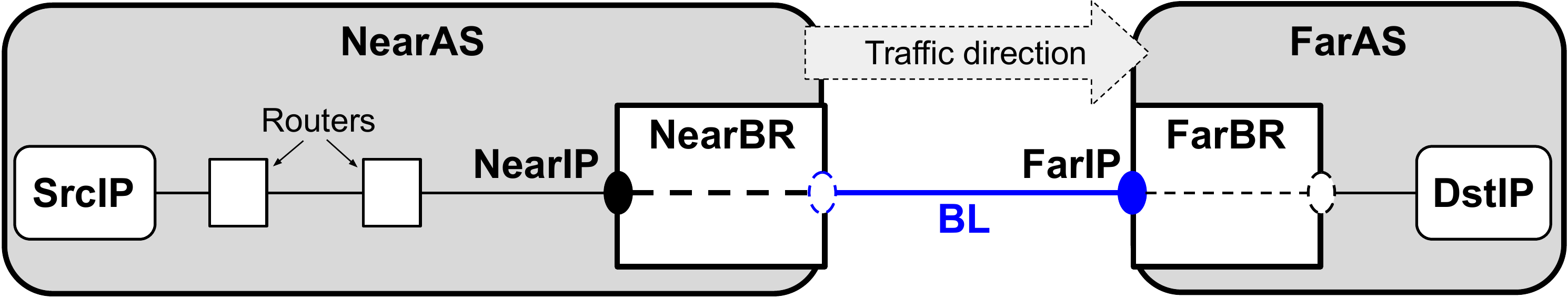}
    \label{fig:NR}}
    \subfigure[ Multipath routing ]{
    \includegraphics[width = 0.8\textwidth]{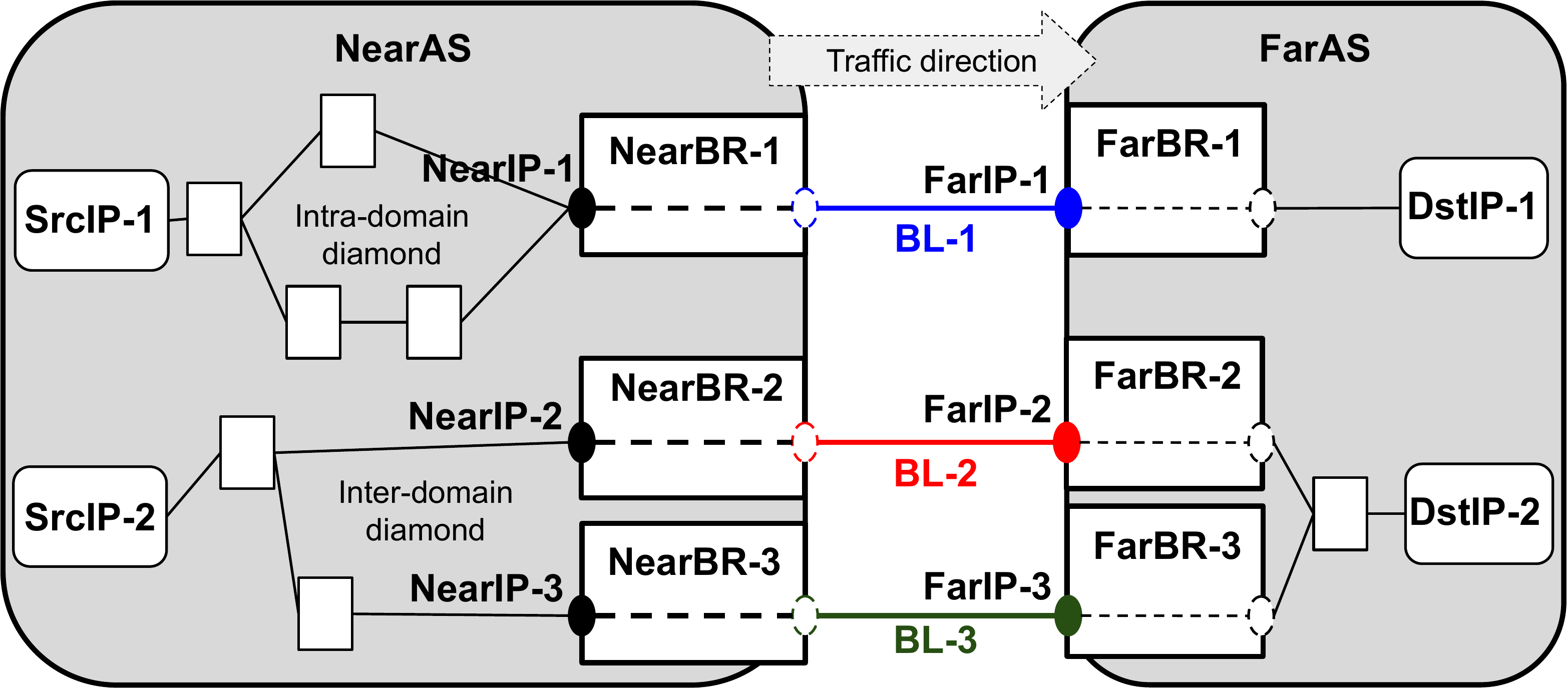}
    \label{fig:MR}}
    \subfigure[BGP-Multipath routing]{
    \includegraphics[width = 0.8\textwidth]{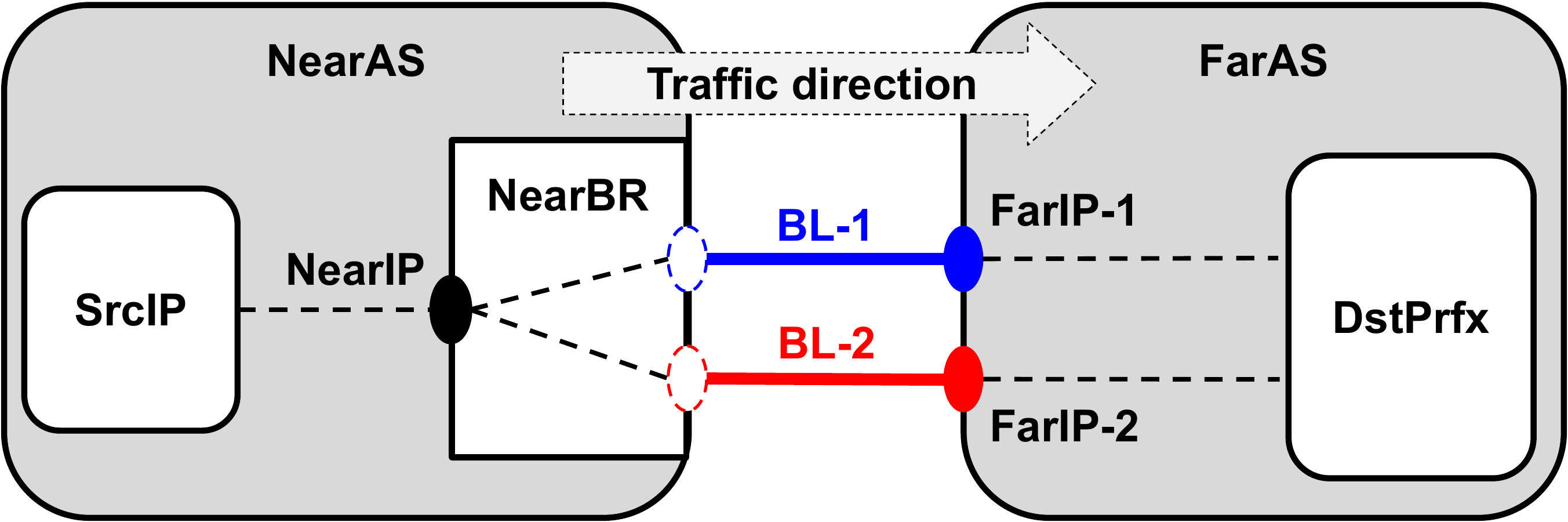}
    \label{fig:BGP-M}} 
\caption{Examples of Best-path routing,  Multipath routing  and BGP-Multipath (BGP-M) between two neighbouring ASes. 
See Table~\ref{tab:Notations} for description of notations.    %
 (a)~Best-path routing, where a single best path is chosen for routing.       %
 %
 (b)~Multipath routing, where an intra-domain router divides traffic to a \emph{DstIP} onto two different paths. 
 If the paths merge within the same AS, they form a so-called  intra-domain `diamond'; if they cross AS borders, they  form  an inter-domain diamond. 
 (c)~BGP-Multipath (BGP-M) routing, where a border router shares traffic to  a \emph{DstPrfx} on two inter-domain border links.
   }
   \label{fig:Illustration}
\end{figure}

\begin{table}[t]
\centering
\caption{BGP Best Path Selection Algorithm  
}
\begin{tabular}{c|l|l}
\hline 
{\bf Priority} & {\bf Attribute}    & {\bf Best path selection rules}  \\
\hline
1        & LocPref          & Highest local preference \\
\hline
2        & AS path          & Shortest AS path \\
\hline
3        & Origin           & \begin{tabular}[c]{@{}l@{}}Lowest origin type\\ (IGP < EGP < INCOMPLETE)\end{tabular} \\
\hline
4        & MED              & Lowest MED (Multi Exit Discriminator). \\
\hline
5        & eBGP/iBGP        & Prefer eBGP over iBGP paths. \\
\hline
6        & IGP metric       & Lowest IGP metric \\
\hline
7        & Router ID        & Lowest  router ID \\
\hline                                                                                                             \end{tabular}
\label{tab:best-path-selection}
\end{table}

\section{Background}
\label{Sec:Background}

\subsection{Border Router and Border Link}
Although network operators of ASes can  use various intra-domain  protocols for routing within  boundary of their own networks, BGP~\cite{RFC4271} is the default inter-domain protocol used universally for routing among ASes throughout the  global Internet. BGP is policy-based and allows a lot of flexibility in implementing routing policies.

A border router, also called BGP border router or AS border router, is located at the boundary of an AS with at least one interface connecting to an intra-domain router and at least one interface connecting to a border router in a neighbour AS. 
A border router is implemented with BGP. 
It can establish and maintain BGP sessions to exchange routing information with other ASes via BGP messages, and then update its routing table according to the network operator's policy configurations.

{
When two ASes have established BGP sessions via border routers, they peer with each other and are called two {\em peering ASes}.
Two peering ASes are neighbouring ASes if they are connected directly, via  physical links or an IXP (for reduced latency and improved routing performance), and they are called {\em neighbour ASes} to each other. 
Otherwise, the two peering ASes are called {\em remote ASes} to each other (for monetary savings and increased connectivity \cite{Giotsas2021TON}}.
An AS can deploy BGP-M to both neighbour ASes and remote ASes.
This paper focuses on BGP-M deployed by an AS to its neighbour ASes which are more common.

A border link  is a physical IP-level inter-domain link connecting border routers of two neighbouring ASes.
As illustrated in Figure~\ref{fig:NR}, depending on traffic direction, a border link   starts from an egress interface of a border router of the nearside AS,
and ends at an ingress interface of a border router of the farside AS.
Since the egress interface of a router is invisible in traceroute measurement,   a border link  is usually denoted by the ingress interfaces of the two border routers{, i.e.~{\em{NearIP}} on {\em{NearBR}} and {\em{FarIP}} on {\em{FarBR}},} which can be identified as two consecutive IP addresses on a traceroute path that are mapped to two different ASes.


\subsection{Best-path Routing}

By default, if a border router  receives advertisements of different routes to a destination prefix, it should select the best path by considering a series of BGP attributes in order of their priority as shown in Table~\ref{tab:best-path-selection}, where Router ID  is only used as  a last-resort tie-breaker if all other attributes have equal values~\cite{RFC4271}.

Until recently, it was expected that there   should normally be a single valid IP-level routing path from a source IP address to a destination IP address (see Figure~\ref{fig:NR}).  
When multiple paths were observed, they were considered  as anomalies, possibly due to routing table misconfiguration~\cite{Javed2013SIGCOMM}, link failures~\cite{Comarela2013IMC, Fanou2015PAM, Paxson1997TON} or 
change of routing paths~\cite{Medem2012INFOCOM, Rimondini2014PAM, Ahmed2015LCN, Cunha2014TON,  Wassermann2017BigDAMA}.

\subsection{Multipath Routing}
\label{Sec:BG_MR}

In recent years, network operators utilised a traffic engineering technology called the {\em multipath routing}, deployed at intra-domain routers within an AS, to  enable multiple IP-level routing paths to a destination (see Figure \ref{fig:MR}). 
These multiple  routes are legitimate, lasting routes. They are used concurrently to balance traffic load in order to achieve improved routing performance and resilience~\cite{Qadir2015IEEECST, Singh2015IEEECST}.

Researchers identified multipath routing from traceroute data 
\cite{Augustin2007E2EMON, Veitch2009INFOCOM,  Augustin2011TON, Vermeulen2018IMC, Vermeulen2020NSDI,  Almeida2020INFOCOM} using Paris traceroute~\cite{Augustin2006IMC} and Multipath Detection Algorithm (MDA)~\cite{Augustin2007E2EMON}. 
The research focus was on the link and node discovery and the topological characteristics of diamonds (see Figure \ref{fig:MR}) or load balancers.

\section{BGP-Multipath}
\label{Sec:BGP-M}

BGP-Multipath (BGP-M) is a load balancing technique deployed at a border router to share traffic load to a destination prefix on different border links using the ECMP function. 

\subsection{Equal-Cost Multi-Path (ECMP)}

Routers produced by most major vendors, such as Juniper, Cisco and Huawei~\cite{juniper-mbgp, CISCO, huawei-mbgp}, have already supported the ECMP function.  
They allow routers to install multiple internal or external BGP paths, called iBGP-Multipath and eBGP-Multipath, respectively.  
This function is called `BGP-Multipath' by Juniper and  Cisco; or `BGP Load Balancing' by Huawei.

\subsection{Deployment of BGP-M}
As shown in Figure \ref{fig:BGP-M}, in order for a network operator ({\em NearAS}) to deploy BGP-M at a border router ({\em NearBR}), the following conditions must be satisfied. 
(1) {\em NearBR} supports the ECMP function; (2) {\em NearBR} has multiple border links   connecting to border router(s) of a same neighbour AS ({\em FarAS}), either directly or via an IXP;
(3) {\em NearBR} has learned from the neighbour AS  multiple routes via different border links, to a given destination prefix ({\em DstPrfx}); and (4)   the multiple routes have equal values for the first 6 attributes in Table\,\ref{tab:best-path-selection}.

If the above conditions are met (i.e.\,the routes learned over different paths are considered sufficiently equal), the network operator can deploy BGP-M at the border router by installing  the multiple routes   in the routing table such that the border router is configured to use these paths concurrently. 
Because all the relevant BGP attributes for the routes over different paths are the same and the border router still announces one route as the best route, there is no impact on BGP loop detection or other BGP processing~\cite{Valera2011MBGP}.  

{Note that the multiple IP-level paths used in {the deployment of BGP-M} always follow the {same} AS-level path.
We used the terms  `Multipath BGP' or `M-BGP' in our preliminary works~\cite{Li2020TMA, Li2021GI}. Since then we have changed to the terms `BGP-Multipath' or `BGP-M' to avoid confusion with `Multi-path BGP' in ~\cite{Fujinoki2008ICON, Valera2011MBGP} and `Multipath BGP' in~\cite{Beijnum2009INFOCOM} that are relevant to {multiple} AS-level paths.
}

\subsection{Limited Documentation on BGP-M}

To a large extent, BGP-M remains an obscure technique  
because there are only a small number of documents related to BGP-M in literature. 

\subsubsection{Router vendor documents} Major router vendors, like Juniper, Cisco and Huawei~\cite{juniper-mbgp, CISCO, huawei-mbgp}, provided technical documentations on the  ECMP function that underlies the BGP-M.

\subsubsection{RFC2992 on Analysis of an Equal-Cost Multi-Path Algorithm (2000)~\cite{RFC2992}}  This Request for Comments (RFC) introduced a hash-threshold method for routers to choose a next-hop (path) from equal-cost multiple paths, which is used for {the deployment of BGP-M}.   

\subsubsection{IETF Draft on Equal-Cost Multipath Considerations for BGP (2019)~\cite{IETF-ECMP}} This Internet-Draft by the Network Working Group of the Internet Engineering Task Force (IETF)   described the application of  ECMP  in various scenarios. This  is perhaps the most relevant document on BGP-M. 

\subsubsection{Research Publications}
Valera {\it et al.}~\cite{Valera2011MBGP} briefly discussed the concept of BGP router using multiple equal-cost paths concurrently.
Mok {\it et al.}~\cite{Mok2018PAM} studied YouTube's load balancing behaviour on  inter-domain border links.
Augustin {\it et al.}~\cite{Augustin2011TON} and Almeida {\it et al.} \cite{Almeida2017PAM} mentioned the possibility of multipath routing based on ECMP.

\subsection{Challenges in Discovering BGP-M from Traceroute Data}
\label{Sec:Inference_Challenges}

{

So far there is no dedicated dataset for BGP-M. 
In the past, traceroute data were used to study  multipath routing deployed at intra-domain routers~\cite{Augustin2011TON, Almeida2017PAM, Vermeulen2020NSDI, Almeida2020INFOCOM}, where specific traceroute tools were designed and deployed and large amounts of data were collected.
In theory, traceroute with UDP packets has the potential to discover BGP-M deployed at border routers, but there are a number of challenges. 

One challenge is that we will need to design  a traceroute probe specially customised for discovering BGP-M. 
Then, without any prior knowledge, we will have to deploy the traceroute probe in as many ASes as possible; and from each probe, we will have to run traceroute to as many different destination prefixes in as many  other ASes as possible. 

The largest challenge, however, is the lack of sound tools or datasets for  IP-to-AS mapping and AS border mapping -- despite more than a decade of research effort. 
If we use traceroute data to discover BGP-M, we must be able  to accurately determine the border of an AS on a traceroute path, so that we can credibly identify border router and border links. 
A recent study~\cite{Yeganeh2019IMC} shows that existing efforts on IP-to-AS mapping and AS border mapping~\cite{Huffaker2010PAM,Pansiot2010PAM,Nur2018Comput.Netw.,Giotsas2015CoNEXT,mi2,Luckie2016IMC,Marder2016IMC,Marder2018IMC}  still cannot avoid erroneous results.

}

\section{Inference of {the Deployment of BGP-M} \\ from Looking Glass (LG) Data}
\label{Sec:Inference}

Here we introduce our effort in inferring {the deployment of BGP-M} in the Internet based on query data from Looking Glass (LG) servers. 
This method has a number of advantages: it is relatively easy to conduct and analyse, it reveals a rich set of information, and most importantly, it is accurate and credible such that our result can be considered as ground-truth.  

\subsection{Notation of a BGP-M Case}
\label{Sec:Inference_Definition}
 
Table~\ref{tab:Notations} lists the notations used in this paper. We used a 4-tuple, {\em <NearAS, NearBR, FarAS, DstPrfx>},  to denote a unique   case of BGP-M deployment, or   BGP-M case. 
We chose these 4 parameters because (1) these values can be observed and confirmed in the routing information retrieved from LG server, and (2) they  defined the owner ({\em NearAS}) and location ({\em NearBR}) of a BGP-M case as well as the destination ({\em DstPrfx}) and the neighbour AS ({\em FarAS}) from which the multiple paths were learned from. 
A BGP-M case is valid for traffic from any source and therefore is irrelevant to {\em{SrcIP}}. 
Border links used by a BGP-M case are between {\em NearAS} and {\em FarAS},  and their {\em FarIPs} are listed  in the response from LG  server\footnote{As we will show later,  border links may change from time to time.}.

A {\em{NearAS}} can deploy BGP-M   at different {\em{NearBRs}} for the same {\em{DstPrfx}}; or   at the same {\em{NearBR}} for different {\em{DstPrfxes}}. 
All these are considered as different cases of BGP-M deployment as they have different tuples. 

For convenience, when we studied a BGP-M case,  we only considered traffic routing between two neighbouring ASes, i.e.\,traffic started in {\em{NearAS}} and ended in {\em{FarAS}}. 
In the real Internet, {\em{SrcIP}} can be outside   {\em{NearAS}}  and {\em{DstPrfx}} can be outside  {\em{FarAS}} -- indeed they can be anywhere on the Internet as long as the traffic {for the {\em DstPrfx}} traverses through {\em{NearAS}} and {\em{FarAS}} via {\em{NearBR}} {in a BGP-M case}. 
If {\em{NearAS}} and {\em{FarAS}} are indirectly connected at an IXP, the BGP-M tuple does not need to include the IXP {because IXPs are `transparent' in BGP routing (and usually considered as a part of the {\em{FarAS}})~\cite{RFC7947}. 
In other words, the existence of IXP does not affect the function and the deployment of BGP-M. 
}

\subsection{Our Inference Method} 
\label{Sec:Inference_LG}

\subsubsection{LG servers}

A Looking Glass (LG) server provides Web-based interfaces at one or more border routers to allow non-privileged execution of network commands (e.g.\,traceroute, ping, and BGP). 
These commands provide direct access to the BGP configuration and routing tables of border routers beyond what is propagated through BGP updates collected by RouteViews~\cite{RouteViews} and RIPE RIS~\cite{RIPERIS}. 
LG server data have been  used to study the Internet topology and path diversity~\cite{Chang2004Comput.Netw., Zhang2005CCR, Han2006TDS, Khan2013IMC, Holbert2015TNSM}.  
Recently the Periscope platform~\cite{Giotsas2016PAM} was proposed to unify LG servers with publicly accessible querying API and to support on-demand measurements. 

We proposed to infer {the deployment of BGP-M} by querying LG servers because they can provide non-transitive BGP attributes containing direct and reliable information on {the deployment of BGP-M}. 

\begin{table}[]
    \centering
    \caption{LG servers in the Top-200  ASes as ranked by CAIDA~\cite{caida_asrank}}
    \label{tab:LGServers}
    \begin{tabular}{l|r|r|r|r}
    \hline
    
        \multirow{2}{*}{  }     & \multicolumn{4}{c}{\# of ASes  in each group in  Top-200 } \\ 
        \cline {2-5}
            &  1--10 &  11--50 &  51--200 & Total \\  \hline
        With known LG URL &   10 & 31 & 82 & 123  \\ 
        With accessible URL &  8 & 20 & 62 & 90 \\
        Support \texttt{routes} command   & 4 & 11 & 36 & 51\\
        With identified BGP-M cases & 1 & 3 & 0   & 4 \\
        \hline
         
    \end{tabular}

\end{table}

\subsubsection{List of ASes with LG servers}
In our work, we firstly compiled a list of ASes with LG servers from a number of data sources including    BGP Looking Glass Database~\cite{BGPLGDatabase}, PeeringDB~\cite{PeeringDBAPI, PeeringDB}, and traceroute.org~\cite{traceroute}. 
The list contained 2,709 AS numbers (ASNs). 
Table \ref{tab:LGServers} lists the number of ASes with LG servers in the Top 200 ASes as ranked by CAIDA~\cite{caida_asrank}, where in total 51 ASes had accessible LG servers and supported the \texttt{routes} command (e.g.\,{\tt show ip bgp routes detail <IP address>}) which is needed for our inference~\cite{Li2020TMA}.
 
For clarity, in the following we call a border router to which we send LG enquiry a `nearside border router' ({\em NearBR}); and an AS that owns and manages the nearside border router a `nearside AS' ({\em NearAS}). 

\begin{figure}
    \centering
    
    \includegraphics[width = 0.8\textwidth]{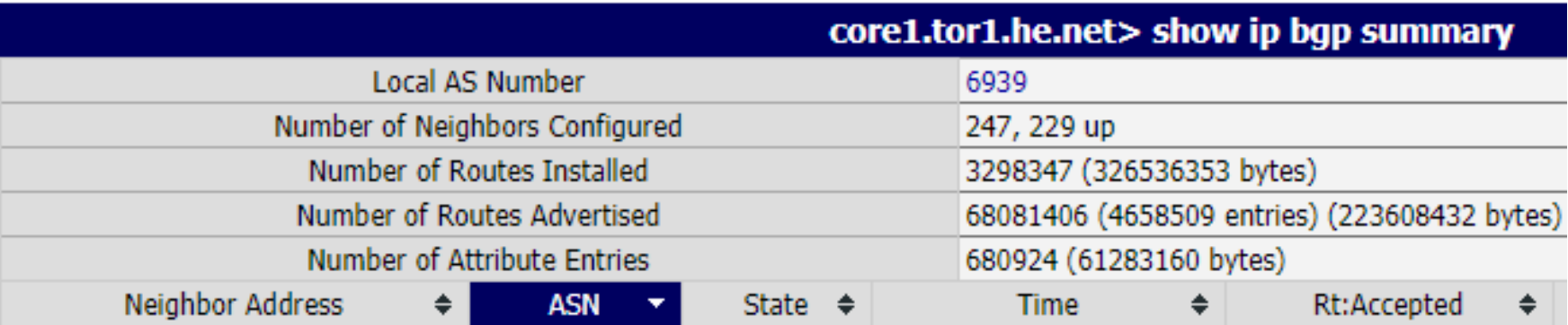}
    \includegraphics[width = 0.8\textwidth]{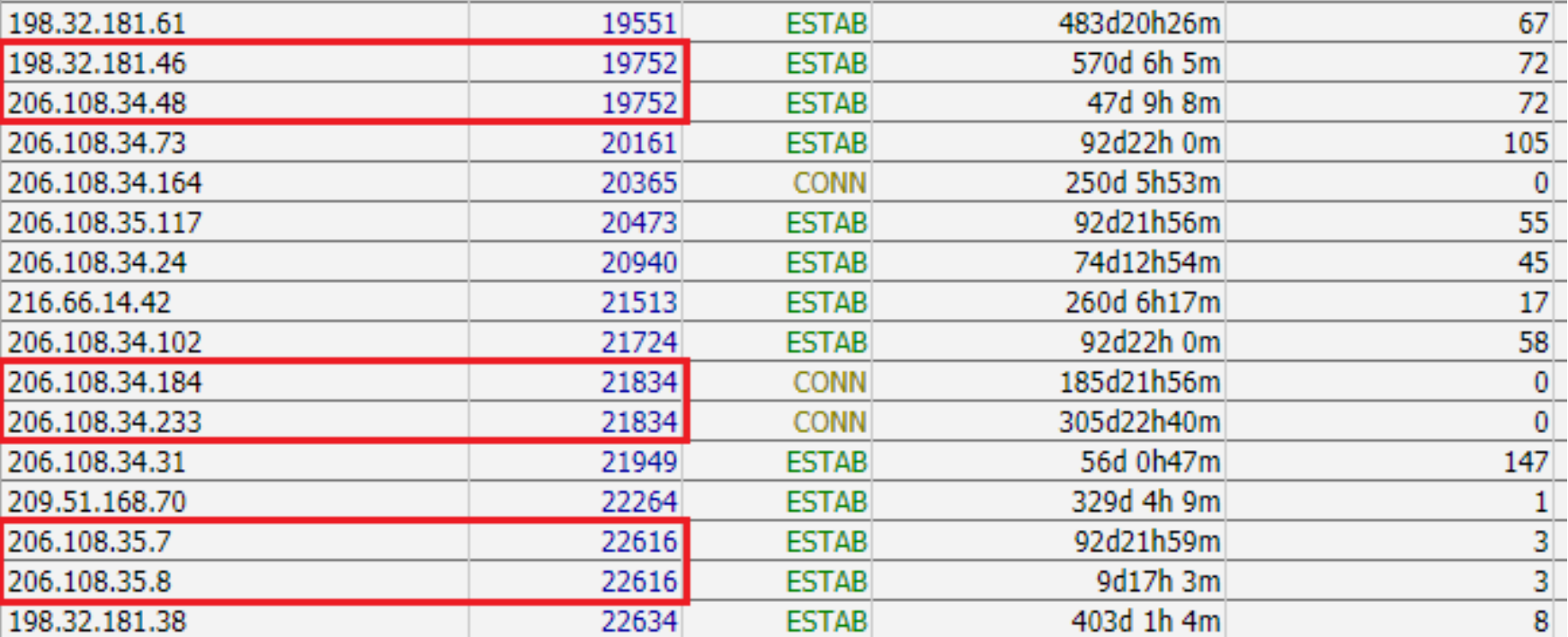}
    \caption{An example of LG response from  border router {\tt core1.tor1.he.net}  of  Hurricane Electric (AS6939) to the {\tt summary} command. Each red box   highlights a neighbour AS with multiple border links connected to the border router.}
    \label{fig:AS6939BGPSummary}
\end{figure}

\begin{figure*}[!t]
    \centering
    
     \includegraphics[width= 1.3 \textwidth]{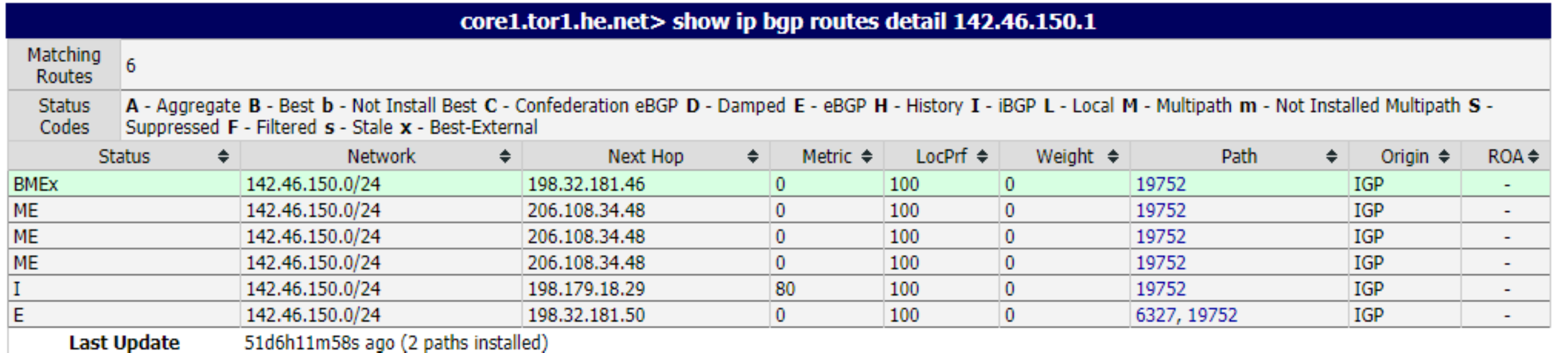}
    \caption{An example of LG response to the {\tt routes} command ({\tt show ip bgp routes detail <IP address>}), from the border router {\tt core1.tor1.he.net} of Hurricane Electric.}
    \label{fig:AS6939BGPRoute}
\end{figure*}

\begin{landscape}
\begin{table*}[!t]
    \centering  
    \caption{ASes with  {the Deployment of BGP-M} in the Internet}
    \label{tab:AS-BGPM}
    \begin{tabular}{r|l|c||r|c||r|r|r||r|r|r}

    \hline
       AS &  AS  &  AS  & \multicolumn{2}{c||}{\# of BGP-M cases}  & \multicolumn{3}{c||}{\# of neighbour ASes} & \multicolumn{3}{c}{\# of nearside border routers}\\ \cline{4-11}
    number & name & rank &  total & IXP~/~Direct~/~Hybrid   & total  & with BGP-M & ratio  & total & with BGP-M & ratio\\ \hline
    
     \multicolumn{11}{c}{IPv4}  \\  \hline
    6939    & Hurricane Electric    & 7     & 1,088 & 1,006/68/14   & 5,868 & 611 & 10.4\%    & 112   & 69    & ~61.6\%   \\
    9002    & RETN                  & 13    & 155   & 87/65/3       & 1,547 & 108 & ~7.0\%    & 130   & 51    & ~39.2\%   \\
    3216    & PJSC VimpelCom        & 25    & 2     & 0/2/0         & 770   &  2   & ~0.3\%    & 16    & 2     & ~12.5\%   \\
    20764   & CJSC RASCOM           & 30    & 27    & 23/4/0        & 858   & 23  & ~2.7\%    & 27    & 6     & ~22.2\%   \\
    12303   & ISZT                  & stub  & 2     & 2/0/0         & 59    &  2   & ~3.4\%    & 2     & 1     & ~50.0\%   \\ 
    22691   & ISPnet                & stub  & 3     & 0/3/0         & 24    &  3   & 12.5\%    & 7     & 1     & ~14.3\%   \\
    48972   & BetterBe              & stub  & 2     & 2/0/0         & 9     &  1   & 11.1\%    & 4     & 2     & ~50.0\%   \\ 
    52201   & {TCTEL}                 & stub  & 1     & 0/1/0         & 11    &  1   & ~9.1\%    & 1     & 1     & 100.0\%   \\
    131713  & {IDNIC}                 & stub  & 1     & 1/0/0         & 10    &  1   & 10.0\%    & 5     & 1     & ~20.0\%    \\
    196965  & TechCom               & stub  & 24    & 24/0/0        & 36    & 15  & 41.7\%    & 2     & 2     & 100.0\%   \\
    328112  & {LBSD}                & stub  & 13    & 0/2/11        & 29    & 13  & 44.9\%    & 2     & 1     & ~50.0\%    \\ \hline
    
     \multicolumn{11}{c}{IPv6}  \\  \hline
    6939    & Hurricane Electric    & 7     & 300   & 266/14/20     & 3,880 & 146 & ~3.8\%    & 112   & 35    & ~31.3\%   \\
    9002    & RETN                  & 13    & 45    & 25/18/2       & 926   & 23  & ~2.5\%    & 130   & 24    & ~18.5\%   \\
    8647    & {AS-T2012}              & stub  & 2     & 2/0/0         & 46    &  2   & ~4.3\%    & 1     & 1     & 100.0\%      \\
    48972   & BetterBe              & stub  & 2     & 2/0/0         & 6     &  1   & 16.7\%    & 4     & 2     & ~50.0\%   \\
    131713  & {IDNIC}                 & stub  & 1     & 1/0/0         & 5     &  1   & 20.0\%    & 5     & 1     & ~20.0\%    \\
    328112  & {LBSD}                & stub  & 6    & 6/0/0          & 28    &  6   & 21.4\%    & 2     & 1     & ~50.0\%    \\ \hline
    \multicolumn{11}{l}{IDNIC: IDNIC-SPICELINK-AS-ID} \\
    \multicolumn{11}{l}{LBSD: Linux-Based-Systems-Design-AS} \\
    \end{tabular}
\end{table*}
\end{landscape}

\subsubsection{List of neighbour ASes}
\label{Sec:Inference_neighbourAS}
For each {\em NearBR}, we obtained a list of neighbour ASes that were connected to the {\em NearAS} at the {\em NearBR}. 

Some ASes provided both the \texttt{routes} command and the \texttt{summary} command (e.g.\,{\tt show ip bgp summary}). 
The \texttt{summary} command allowed us to not only find the neighbour ASes, but also identify those neighbour ASes connected to the {\em NearBR} via multiple border links.
Figure \ref{fig:AS6939BGPSummary} is an example table returned by  the \texttt{summary} command from {\tt core1.tor1.he.net} ({\tt tor1}), a border router of Hurricane Electric. 
The table lists the ASNs of the BGP neighbours and the IP addresses of the interfaces through which the BGP sessions are established. 
Those neighbour ASes highlighted in red boxes were connected to {\tt tor1} via multiple neighbour addresses, i.e.\,via multiple border links,  and therefore were potential candidates for {the deployment of BGP-M}.  

{For ASes that did not provide the \texttt{summary} command,  we are unable to directly obtain their neighbour ASes connected to border routers via multiple border links. 
Therefore, we firstly obtained the AS paths from BGP RIB entries provided by RouteViews~\cite{RouteViews}.
Then for each {\em NearAS}, we extracted its neighbour ASes as those next to it in any RIB AS path.
Afterwards, we query all the extracted neighbour ASes, which requires more  analysis in our inference.}

\subsubsection{Retrieving routing table}
\label{Sec:Inference_Identification}

For each {\em NearBR}, we retrieved its routing table information using  the  \texttt{routes} command, e.g. {\tt show ip bgp routes detail <IP address>}.  
Queries to any IP address in a prefix should return the same routing table. Hence, we only queried one IP address in each prefix.
Thus we set the parameter {\tt IP address}  as {\tt X.Y.Z.1} for IPv4 (or {\tt X:Y:Z::1} for IPv6) for each prefix in a neighbour AS. 
We obtained the full list of prefixes in each neighbour AS from data provided by RouteViews~\cite{RouteViews}.
For simplicity, we only considered   /24  prefixes for IPv4 and /48 prefixes for IPv6, which accounted for 57.5\% and 45.8\% in the RouteViews data for IPv4 and IPv6, respectively.

\subsubsection{Identifying {BGP-M Cases}} 

Figure~\ref{fig:AS6939BGPRoute} shows an example response to the command from the border router {\tt tor1} of  Hurricane Electric.
The router has learned and installed two paths, via different next hops 198.32.181.46 and 206.108.34.48 (i.e.\,two border links), towards the same destination prefix (142.46.150.0/24) in the neighbour AS (AS19752). 

Both paths are labelled with status codes of ``M'' and ``E'', meaning they are multipath  learned via eBGP. 
The two paths all have the same values for    attributes of LocPref, AS Path, Origin, Metric `0' (for IGP) and MED (not shown in the figure), suggesting they are equal-cost multiple paths.
This is the ground-truth evidence that Hurricane Electric (AS6939) has deployed BGP-M  at {\tt tor1} to a destination prefix in the neighbour AS19752.
This BGP-M case is denoted as <AS6939, {\tt tor1}, AS19752, 142.46.150.0/24>.

In this study, we did not query all prefixes in a neighbour AS because we aimed to reduce the total number of queries to LG servers, which often set a cap on the number or frequency of queries from a host. 
If a prefix in a neighbour AS was identified as having BGP-M at the border router, we stopped querying prefixes in that neighbour AS and we continued with another neighbour AS. 
Thus, we may not uncover all BGP-M cases, but we were able to reveal deployment of BGP-M to as many neighbour ASes as possible and each case we inferred was a confirmed ground-truth case. 
In other words, our inference is a conservative, lower bound estimate of the scale of {the deployment of BGP-M} in the Internet. 

If all  prefixes in a neighbour AS were queried and no BGP-M was identified, the query went to another neighbour AS. 
This did not indicate that BGP-M was not deployed at the border router to the neighbour AS, because we only queried   prefixes of size /24 (or /48) in the neighbour AS.

When all the obtained neighbour ASes were queried for a border router, the query went to the next border router. When all the border routers of a {\em NearAS} were queried, the query went to another {\em NearAS}. 
For {\em NearASes} that provided the {{\tt summary}} command, we only queried the neighbour ASes with multiple border links; and for other {\em NearASes}, we queried all of their neighbour ASes.

\subsection{Our Inference Results}\label{Section:OurInferenceResult}

Table \ref{tab:AS-BGPM} summaries our inference results obtained by applying the above method to  all 2,709 ASes  with an LG server. 
It shows that BGP-M has been widely deployed not only by large transit ASes, such as Hurricane Electric, RETN, PJSC VimpelCom and CJSC RASCOM, but also by many stub ASes. 
BGP-M is deployed on both IPv4 and IPv6 Internet.

Note that although we tried to discover as many BGP-M cases as possible, our inference result was far from a complete measurement. 
In fact, only a small portion of ASes provide a LG server, of which only a small portion are publicly accessible and support the {\tt routes} command.  
In addition, as explained above,   we only had limited resource and time to uncover at most one BGP-M case (for one of many destination prefixes) in each neighbour AS. 
Thus, it is highly likely that there are a lot more    BGP-M cases already deployed in the Internet. 
Our inference result provided a lower bound estimation.

\section{Analysis of BGP-M Cases Deployed by Hurricane Electric}
\label{Sec:BGP-MDeployment}

The most notable AS in our inference result is Hurricane Electric (HE, AS6939). 
It is a  Tier-1 network, ranked 7th in the Internet. 
As a major Internet service provider, it had 112 border routers neighbouring with  5,868 ASes on IPv4  in January 2020. 
It is remarkable that it has already extensively implemented {\em at least} 1,088 BGP-M cases to (at least one) prefixes in 611 of its neighbour ASes at 69 border routers.  
{Note that there could be many more cases of BGP-M to be discovered, i.e., those deployed to other prefixes in these neighbour ASes and those deployed to prefixes in remote ASes.}

Of the 1,088 BGP-M cases on IPv4, 911 cases used 2 border links, 92 cases used 3 links, and 85 cases used 4 links. 
Of the 300 BGP-M cases on IPv6,  248 cases used 2 links, 33 cases used 3 links, and 19 cases used 4 links. 
There are much less BGP-M cases on IPv6 than on IPv4, possibly because there is less demand for load balancing on IPv6 than on IPv4.

\subsection{BGP-M Cases via IXP}
\label{Sec:BGP-MDeployment_HE_IXP}

We relied on data from PeeringDB~\cite{PeeringDB} to identify whether a BGP-M case was deployed via IXP or not. 
We compiled a list of IXPs and the prefixes belonging to them. 
If all of the {\em{FarIPs}}  in a case belonged to IXPs, this case was identified as via {\em IXP}; if none of the {\em{FarIPs}}  in a case belonged to IXPs, this case was via {\em Direct} links; otherwise, this case was  via {\em Hybrid} links.  

It is notable that IXPs played a vital role in HE's {the deployment of BGP-M} as they were involved in 92.5\% (i.e.\,1,006 in 1,088) of the IPv4 cases, and 88.9\% (i.e.\,266 in 300) of the IPv6 cases. 

{Because we only considered one data source on IXP data, some cases via IXP might be mis-classified as via direct links or via hybrid links.
Thus, IXPs might be more important in {the deployment of BGP-M} than we observed.}

\begin{figure}
    \centering
    \subfigure[611 neighbour ASes on IPv4]{
    \includegraphics[width = 0.8 \textwidth]{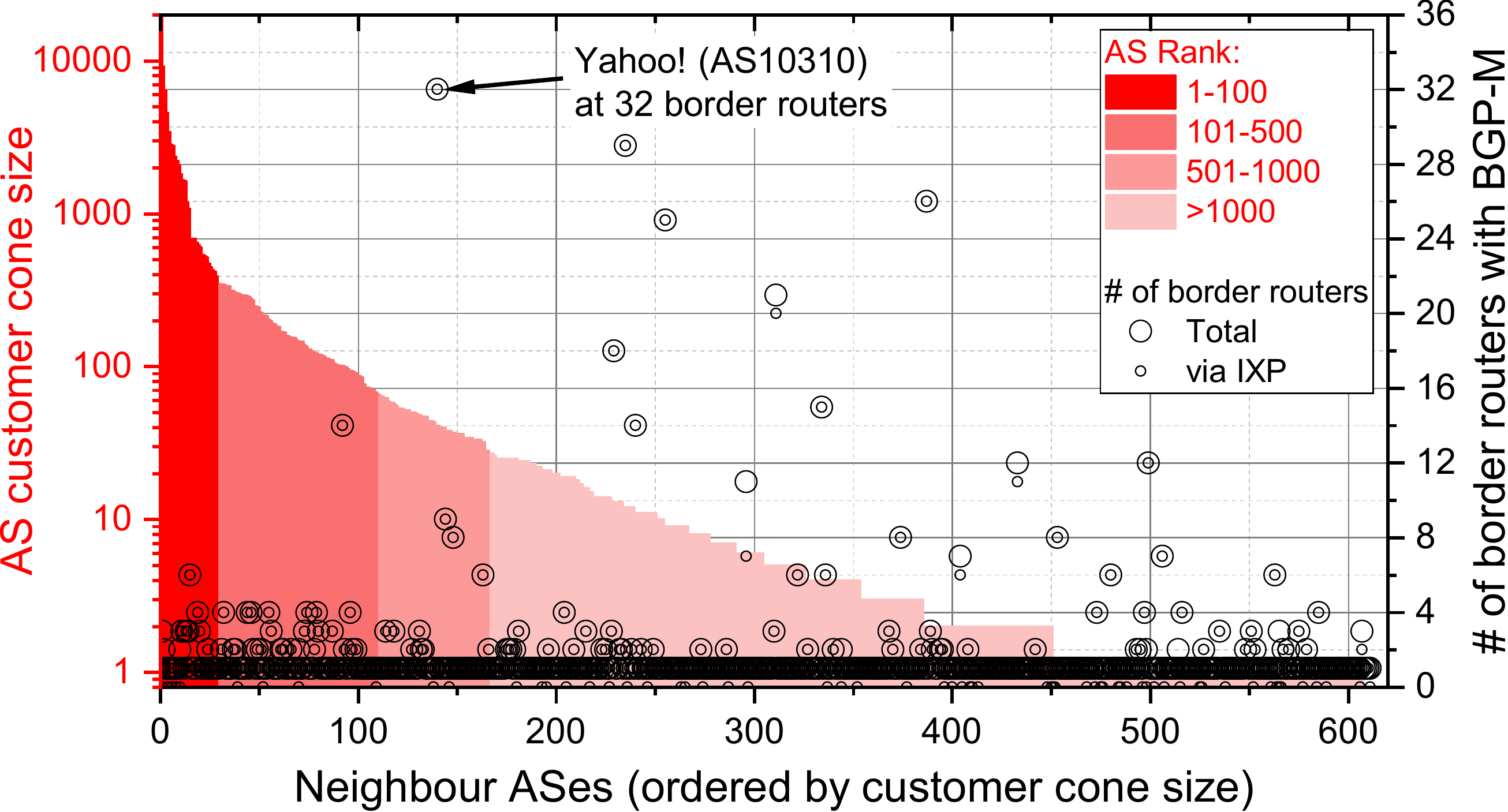}
    \label{fig:NeighbourASes-a}}
    \subfigure[146 neighbour ASes on IPv6]{
    \includegraphics[width = 0.8 \textwidth]{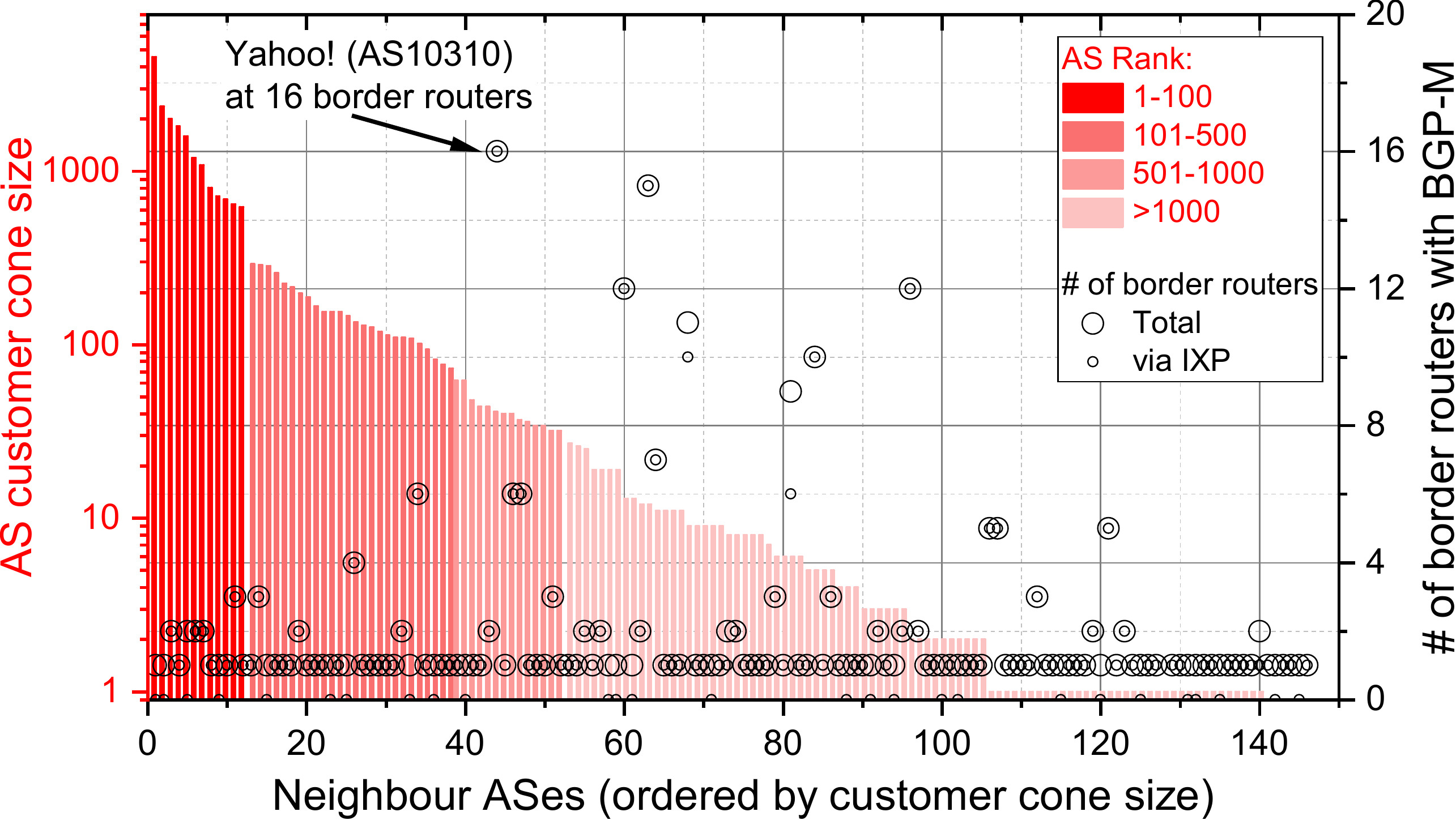}
    \label{fig:NeighbourASes-b}}
    \caption{Hurricane Electric (HE, AS6939)'s neighbour ASes deployed with BGP-M. The neighbour ASes are ordered by their customer cone sizes (y axis on the left in red colour). Also shown is the total number of HE border routers with BGP-M deployment (y axis on the right in black colour) to each neighbour AS, and the number of border routers with BGP-M to each neighbour AS via IXP.}
    \label{fig:NeighbourASes}
\end{figure}

\subsection{Analysis on Neighbour ASes}
\label{Sec:BGP-MDeployment_HE_NeighbourASes}

Figure \ref{fig:NeighbourASes-a} plots the 611 neighbour ASes of HE with at least one BGP-M case on IPv4, ordered by each AS' customer cone size~\cite{caida_asrank, Luckie2013IMC}. 
{
The customer cone of an AS \textit{X} is a set of ASes including (1) \textit{X}'s customer ASes, and (2) \textit{X}'s customer ASes' customer ASes, and so on. The customer cone size of AS \textit{X} is the number of ASes in  \textit{X}' customer cone~\cite{Luckie2013IMC}.
}
%
The plot also shows the total number of HE border routers deployed with BGP-M to a neighbour AS in a large circle, and the number of HE border routers deployed with BGP-M to a neighbour AS via {IXP} in a small circle. 

There are three interesting observations. 
Firstly, Yahoo! (AS10310), a content provider network with customer cone size of 41 and AS rank of 747, was deployed with BGP-M by HE at as many as 32 border routers.   
Secondly, small \& medium ASes (with customer core size < 100) were more likely to be deployed with BGP-M at multiple border routers, suggesting Hurricane Electric has deployed richer and more complex connections to small \& medium ASes than to top-rank ASes. 
Third, IXPs were widely involved in Hurricane Electric's {the deployment of BGP-M}. 
For many neighbour ASes, all of their BGP-M cases were connected through IXP. 
It is very likely that  HE's heavy reliance on IXP is a reason why we observed so many BGP-M cases with small \& medium ASes. 

Table \ref{tab:10ASes-v4} lists the 10 highest ranked neighbour ASes deployed with BGP-M by HE on IPv4 and IPv6. 
As can be seen, these ASes were deployed with BGP-M at only a few ($<=3$)  border routers.
Table \ref{tab:ASeswithMostBGPM} lists the 10 neighbour ASes with the largest numbers of BGP-M cases implemented by HE on IPv4 and IPv6.
Although these ASes are not highly ranked, they are all well-known content provider networks or content delivery networks, and 
{
most of them are among the list of 15 hyper-giant ASes recognised by  B{\"o}ttger {\it et al.}~\cite{hyper-giants}], where hyper-giant ASes are defined as ASes having wide geographical coverage, large port capacity, and large traffic volumes~\cite{hyper-giants}.
}

A comparison between Table \ref{tab:10ASes-v4} and Table \ref{tab:ASeswithMostBGPM}    highlights the difference  between top-rank ASes and hyper-giant ASes (with low ranks) in terms of the requirement for BGP-M. 
BGP-M was more needed and useful  for routing with content providers, where load balancing can be crucial for delivery of   large traffic volume.

Although the amount of BGP-M cases on IPv6 was much lower than IPv4, they  exhibited similar properties, suggesting  HE has applied similar BGP-M policies on IPv4 and IPv6.

\begin{table}[]
    \centering
    \caption{Highest-ranked neighbour ASes deployed with BGP-M by Hurricane Electric
    }
    \label{tab:10ASes-v4}
    \begin{tabular}{c|r|r|l|c}
    \hline

      CAIDA's   &    & Customer   &  & \# of AS6939\\
       AS  & AS &  cone & AS &  border routers  \\ 
     rank & number  &   size &  name &  with BGP-M\\ \hline\hline
     \multicolumn{5}{c}{IPv4}\\
     \hline\hline
       2  & 1299  & 32,929  & Telia Company          & 3 \\ \hline
       9  & 6461  &  9,175  & Zayo Bandwidth         & 2 \\ \hline
       12 & 9002  &  6,374  & RETN                   & 1 \\ \hline
       13 & 4637  &  4,548  & Telstra International  & 1 \\ \hline
       15 & 12389 &  3,425  & PJSC Rostelecom        & 1 \\ \hline
       24 & 7922  &  2,820  & Comcast Cable          & 1 \\ \hline
       25 & 3216  &  2,777  & PJSC VimpelCom         & 1 \\ \hline
       27 & 9498  &  2,361  & Bharti Airtel          & 1 \\ \hline
       29 & 6830  &  2,218  & Liberty Global         & 1 \\ \hline
       30 & 20764 &  2,073  & CJSC RASCOM            & 2 \\ \hline \hline
          \multicolumn{5}{c}{IPv6}\\
     \hline\hline
       15  & 4637  & 4,548  & Telstra               & 1 \\ \hline
       27  & 9498  & 2,361  & Bharti Airtel         & 1 \\ \hline
       32  & 52320 & 2,005  & GlobeNet              & 2 \\ \hline
       36  & 8359  & 1,810  & MTS PJSC              & 1 \\ \hline
       40  & 4826  & 1,593  & Vocus                 & 2 \\ \hline
       48  & 41095 & 1,190  & IPTP LTD              & 2 \\ \hline
       51  & 8220  & 1,083  & COLT                  & 2 \\ \hline
       57  & 4230  &   805  & CLARO S.A.            & 1 \\ \hline
       61  & 4134  &   720  & CHINANET              & 1 \\ \hline
       65  & 5588  &   686  & GTSCE                 & 1 \\ \hline
       \multicolumn{5}{l}{Telstra: Telstra International Limited} \\
       \multicolumn{5}{l}{GlobeNet: GlobeNet Cabos Submarinos Colombia, S.A.S.} \\
       \multicolumn{5}{l}{Vocus: Vocus Communications} \\
       \multicolumn{5}{l}{COLT: COLT Technology Services Group Limited} \\
       \multicolumn{5}{l}{GTSCE: T-Mobile Czech Republic a.s.} \\
       \multicolumn{5}{l}{CHINANET: CHINANET-BACKBONE}
       
    \end{tabular}
\end{table}

\begin{table}[t]
    \centering
    \caption{Ten neighbour ASes of Hurricane Electric (AS6939) with the largest numbers of BGP-M cases    }
    \label{tab:ASeswithMostBGPM}
    \begin{tabular}{l|r|c|r}
    \hline
    AS   & AS     & \# of & AS    \\
    name & number & cases & rank \\
    \hline\hline
    \multicolumn{4}{c}{IPv4}          \\ \hline\hline
    Yahoo!      & 10310 & 32 & 747      \\ \hline
    Cloudflare  & 13335 & 29 & 1845      \\ \hline
    Apple       & 714   & 26 & 6385     \\ \hline
    MicroSoft   & 8075  & 25 & 2288     \\ \hline
    Twitch      & 46489 & 25 & 33522    \\ \hline
    Fastly      & 54113 & 25 & 38523    \\ \hline
    Amazon      & 16509 & 21 & 3560     \\ \hline
    Google      & 15169 & 18 & 1743     \\ \hline
    Twitter     & 13414 & 15 & 4119     \\ \hline
    WoodyNet    & 42    & 14 & 1931     \\ \hline\hline
    \multicolumn{4}{c}{IPv6}            \\ \hline\hline
    Yahoo!      & 10310 & 16 & 747      \\ \hline   
    Cloudflare  & 13335 & 15 & 1845      \\ \hline   
    Apple       & 714   & 12 & 6385     \\ \hline  
    Google      & 15169 & 12 & 1743     \\ \hline  
    MicroSoft   & 8075  & 11 & 2288     \\ \hline  
    Fastly      & 54113 & 11 & 38523    \\ \hline 
    Amazon      & 16509 & 10 & 3560     \\ \hline  
    Verizon     & 15133 & 9  & 3172     \\ \hline  
    WoodyNet    & 42    & 7  & 1931     \\ \hline  
    Limelight   & 22822 & 6  & 344      \\ \hline  
    \end{tabular}
\end{table}

\subsection{Analysis on Border Routers}
\label{Sec:BGP-MDeployment_HE_BR}

Hurricane Electric's LG server~\cite{HELG} covered 112 border routers distributed around the world. 
Table~\ref{tab:BRDistribution} shows that most of its routers were located in North America and Europe and many of them have been implemented with BGP-M. 
The geo-locations of the border routers were directly obtained from their names as given by the LG server. 
Although there were only a few border routers located in Asia and other parts of the world, a large portion of them have been implemented with BGP-M. 

\begin{table}[!t]
    \centering
        \caption{Geographical Distribution of Hurricane Electric's Border Routers.}
    \label{tab:BRDistribution}
    \begin{tabular}{l|r c}
    
    \hline
      &   Number of  & with BGP-M \\  
    & border routers  &   deployment (IPv4, IPv6)\\ \hline
    
    {\bf North America} & {\bf 55~~~~} & {\bf 33, 19} \\ 
        {~~~~United States}  & 47 & 26, 15 \\
        {~~~~Canada}  & 8 & 7, 4\\\hline
        
    {\bf Europe } & {\bf 40~~~~} & {\bf 27, 9} \\ 
        {~~~~Germany} & 5 & 4,  0\\ 
        {~~~~United Kingdom} & 3 & 2, 0 \\
        {~~~~France} & 2 & 2,  0\\
        {~~~~Other} & 30 & 19,  9\\ \hline
        
    {\bf Asia} & {\bf 6} & {\bf 4, 4} \\\hline
    {\bf Other} & {\bf 11} & {\bf 5, 3} \\\hline
    {\bf ~~~~~~~~Total} & {\bf 112} & {\bf 69, 35} \\ \hline
    \end{tabular}
\end{table}

\begin{figure}[!t]
    \centering
    {\includegraphics[width = 0.8\textwidth]{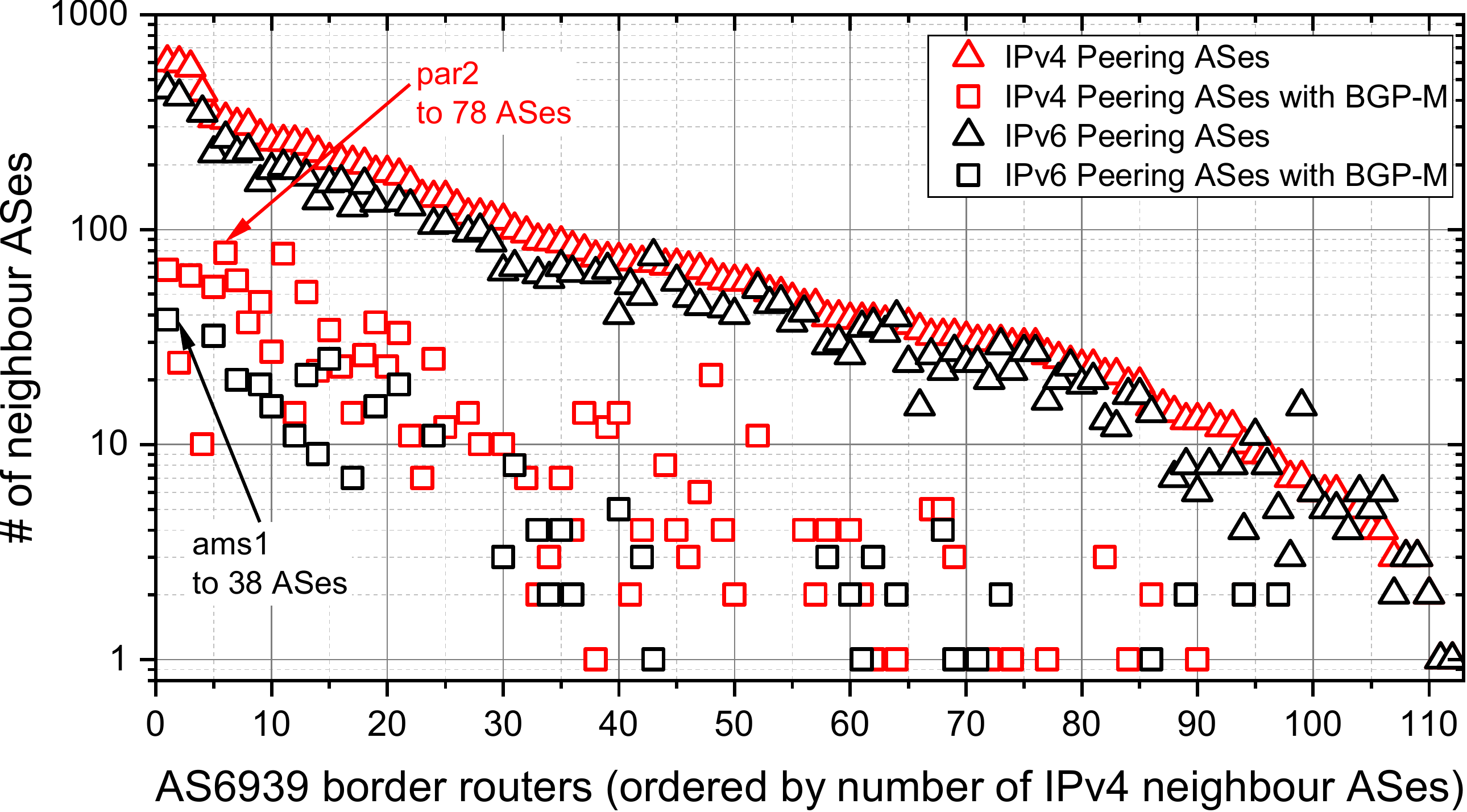}}
    \caption{List of 112 border routers of Hurricane Electric (AS6939). The border routers are ordered by the number of IPv4 neighbour ASes. 
    }
    \label{fig:LG_ASes}
\end{figure}

Figure \ref{fig:LG_ASes} plots the number of neighbour ASes in triangle and the number of neighbour ASes with BGP-M in square at each of HE's 112 border routers on IPv4 and IPv6. 
On IPv4, HE has deployed BGP-M to the largest number (78) of neighbour ASes at the border router {\tt par2}; and on IPv6, the border router {\tt ams1} had BGP-M cases to 38 neighbour ASes.

\begin{figure}[!t]
    \centering
    \subfigure[Number of neighbour ASes  as a function of the number of border routers.]{
    \label{fig:BRandAS-a}
    \includegraphics[width = 0.8\textwidth]{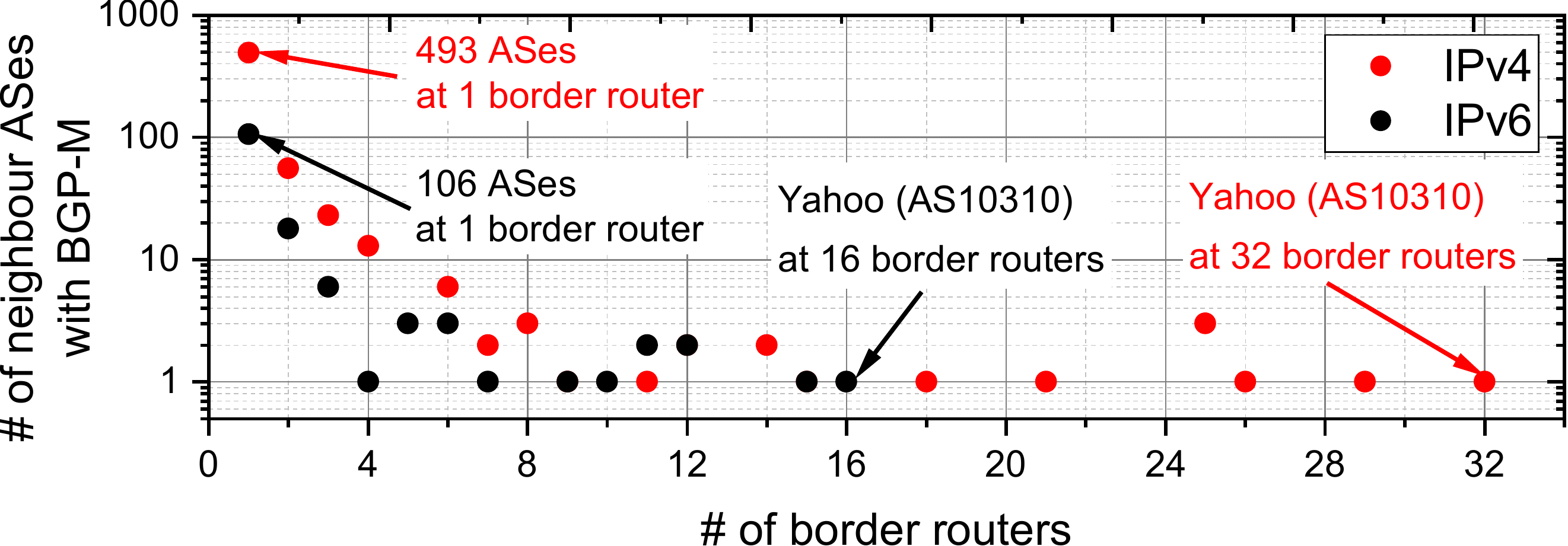}
    }
    \subfigure[Number of border routers as a function of the number of neighbour ASes.]{
    \label{fig:BRandAS-b}
    \includegraphics[width = 0.8\textwidth]{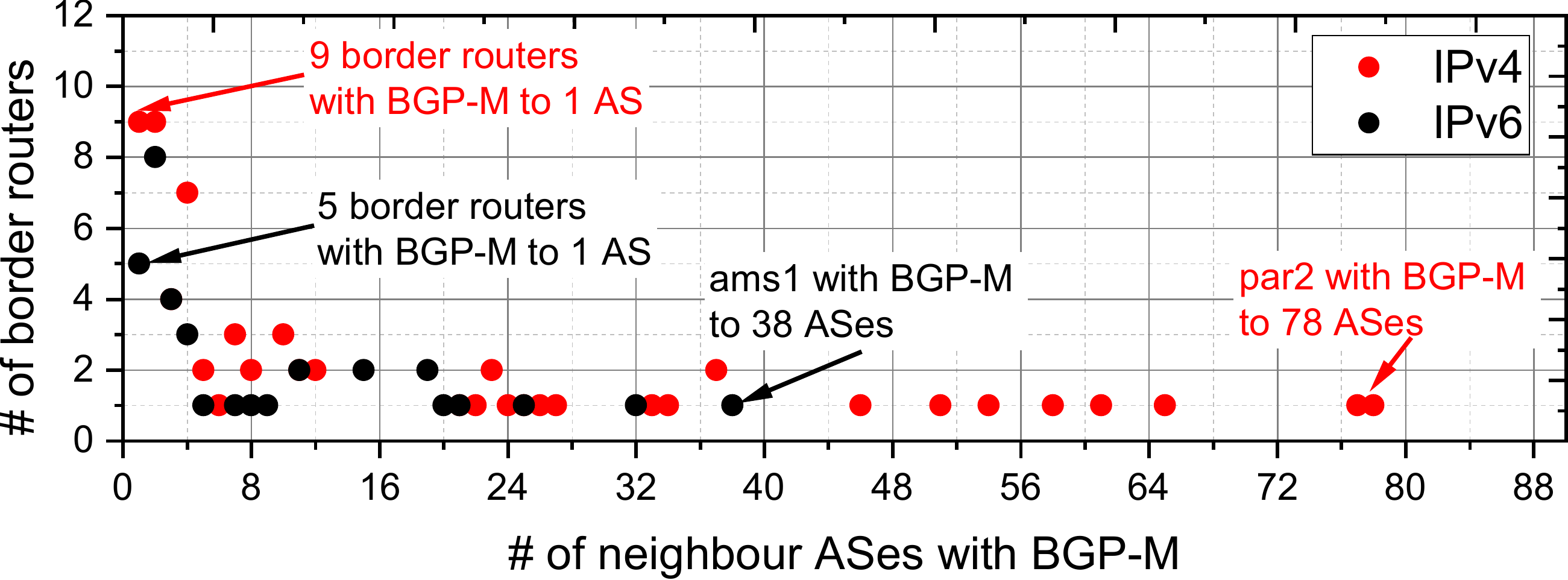}
    }
    \caption{Relation between the number of border routers and the number of neighbour ASes deployed with BGP-M in Hurricane Electric.}
    \label{fig:BRandAS}
\end{figure}

\subsection{Relation between Neighbour ASes and Border Routers}
\label{Sec:BGP-MDeployment_HE_ASvsBR}

Figure \ref{fig:BRandAS} shows the relations between the number of neighbour ASes and the number of border routers deployed with BGP-M by HE. 
Figure \ref{fig:BRandAS-a} shows that HE has deployed BGP-M on IPv4 to 493 neighbour ASes   at only one border router, and BGP-M to the other 118 neighbour ASes  at at least 2 border routers. 
HE has deployed BGP-M with Yahoo! (AS10310) at 32, the largest numbers of, border routers on IPv4,  accounting for 87\% of the border routers connected to Yahoo! Similar observation on IPv6.

Figure \ref{fig:BRandAS-b} shows that on IPv4,   9 of HE's border routers had BGP-M to only one neighbour AS; while other border routers had BGP-M to at least two neighbour ASes. 
The border router {\tt par2} was deployed with BGP-M to the largest number (78) of neighbour ASes. 
It is evident that  BGP-M can be deployed in a flexible way to suit a network's needs.  

\subsection{Summary}

Our inference results from LG data showed that BGP-M has been deployed by both large transit ASes and stub ASes. 
IXPs were widely involved in {the deployment of BGP-M}, suggesting the important role IXPs play in facilitating the deployment of BGP-M.
Moreover, our results revealed that the small \& medium ASes, especially those content provider networks, were more likely to be deployed with BGP-M at multiple border routers, indicating  these ASes' heavy reliance on load balancing to improve their inter-domain traffic delivery.

\section{Study on BGP-M Routing Paths}
\label{Sec:Traceroute}

For a known BGP-M case, we can use traceroute  to reveal  exact details on  how traffic is shared   on border links. 


\subsection{Our Traceroute Probing  }
\label{Sec:Traceroute_ASBorder}

Among existing traceroute projects, including RIPE Atlas~\cite{RIPE2015IPJ}, CAIDA Archipelago (Ark)~\cite{Ark} and iPlane~\cite{Madhyastha2006OSDI}, we found that RIPE Atlas installed publicly accessible traceroute probes in 5 of the 12 ASes where we  identified BGP-M cases (see Table~\ref{tab:AS-BGPM}). 
These 5 ASes were Hurricane Electric,  PJSC VimpelCom, CJSC RASCOM, ISZT  and BetterBe, and they had 3, 3, 4, 2, and 1  RIPE Atlas probes, respectively. 
They had in total  >1,400 BGP-M cases on IPv4 and IPv6.

For each BGP-M case identified as   {\em <NearAS, NearBR, FarAS, DstPrfx>}, we  sent traceroute probings from available RIPE Atlas probes in the {\em NearAS}  to all IP addresses between {\tt X.Y.Z.1} and {\tt X.Y.Z.254}  of {\em{DstPrfx}} on IPv4,  or the first 254 IP addresses between {\tt X:Y:Z::1} and {\tt X:Y:Z::fe} on IPv6.   
We used ICMP packets and UDP packets, Paris traceroute variation 16~\cite{Augustin2006IMC} with default settings on RIPE Atlas, e.g.\,3 packets for probing to each destination IP.

For each BGP-M case, we check whether the traceroute paths sent from a probe to IP addresses in the {\em DstPrfx}  actually traversed the {\em NearBR} where the BGP-M case was deployed.  
If the traceroute paths traverse elsewhere, we discard them.
Below is our procedure.  
\begin{enumerate}[label=(\arabic*)]

    \item Obtain the list of ending points of border links, i.e.\, {\em FarIPs}, {which are given in the routing table returned by the \texttt{routes} command  (see   `Next Hop IPs' in Fig.\,\ref{fig:AS6939BGPRoute}). }  
 
    \item For each traceroute path, check if any of the {\em FarIPs}   appears in the traceroute path. If yes, go to (3); otherwise,  discard this traceroute path.
    
    \item  Use the DNS Chain service provided by RIPEstat Data API~\cite{RIPEstat} to obtain the router name of   the predecessor  IP address of the {\em FarIP}   by using the link of \url{https://stat.ripe.net/data/dns-chain/data.json?resource=<IP address>}. If the router name is {\em NearBR}, finish the process; otherwise, discard this traceroute path. 

\end{enumerate}

Step (3) is necessary because (1) it confirms that the traceroute paths traversed the {\em NearBR} of the BGP-M case under study and (2) it also locates the {\em NearBR} when different BGP-M cases with different {\em NearBRs} share the same {\em FarIPs} of the same {\em FarAS}.

{In this study, we set a standard for traceroute measurement. 
That is, we will only consider traceroute measurement of a BGP-M case if we are able to obtain traceroute data to at least 250 of the 254 IP addresses in the destination prefix and they traverse the relevant {\em NearBR} and {\em BLs}.}

\begin{figure*}[!t]
    \centering
     \subfigure[Topology map.]{
    \includegraphics[width = 0.8\textwidth] {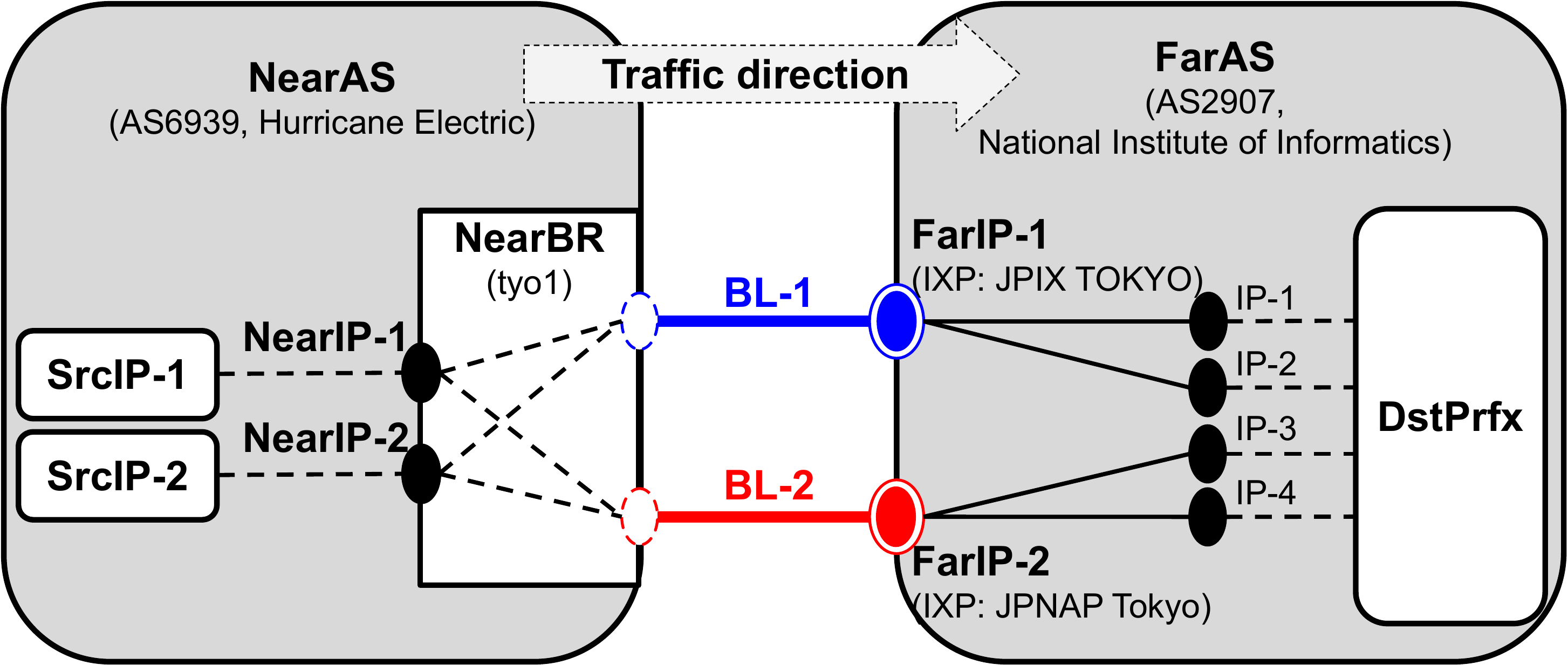}
     \label{fig:IXPCase-a}
     }
     
     \subfigure[Routing map from SrcIP-1 with UDP at {\bf 10:00GMT}]{
    \includegraphics[width = 0.47\textwidth] {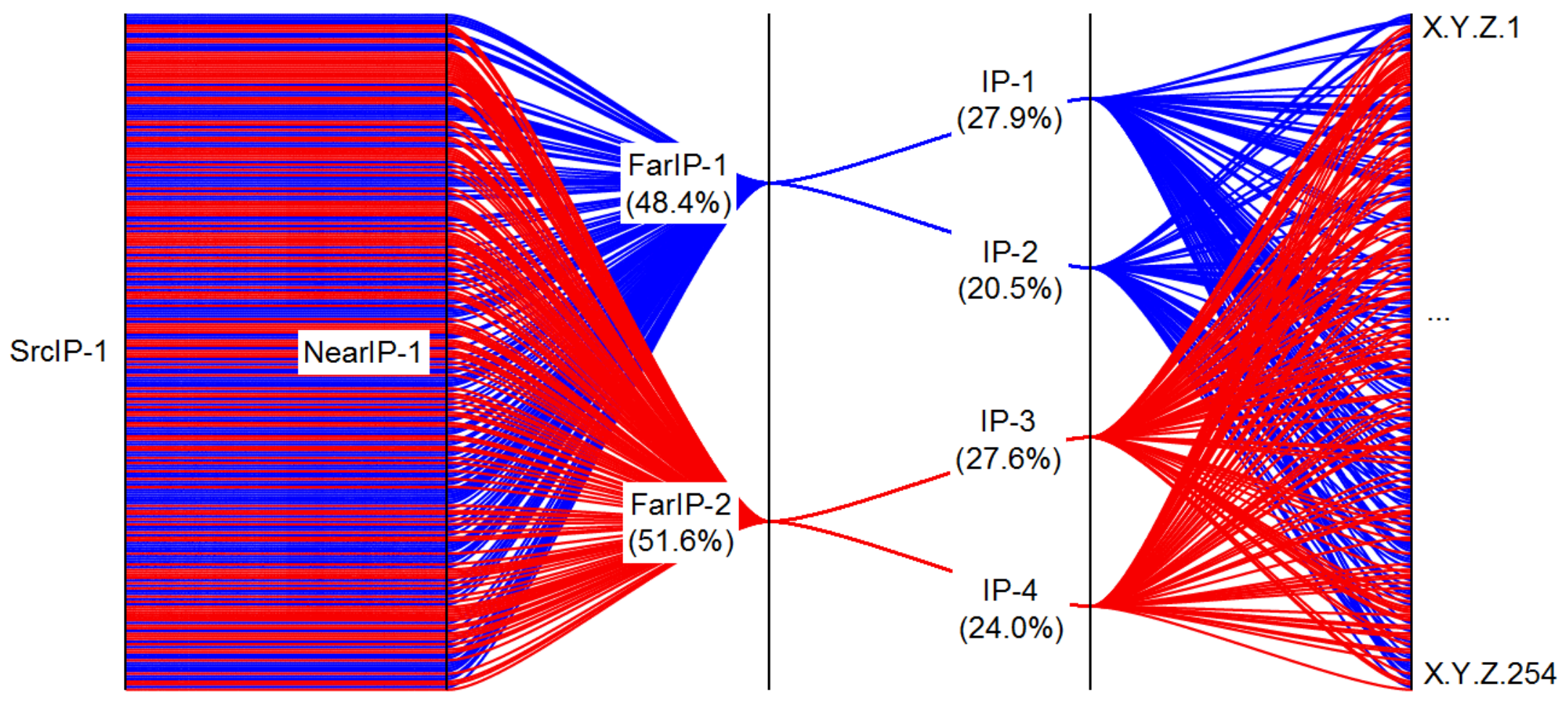}
     \label{fig:IXPCase-b}
     }
    \subfigure[Routing map from SrcIP-1 with UDP at {\bf 10:15GMT}]{
    \includegraphics[width = 0.47\textwidth] {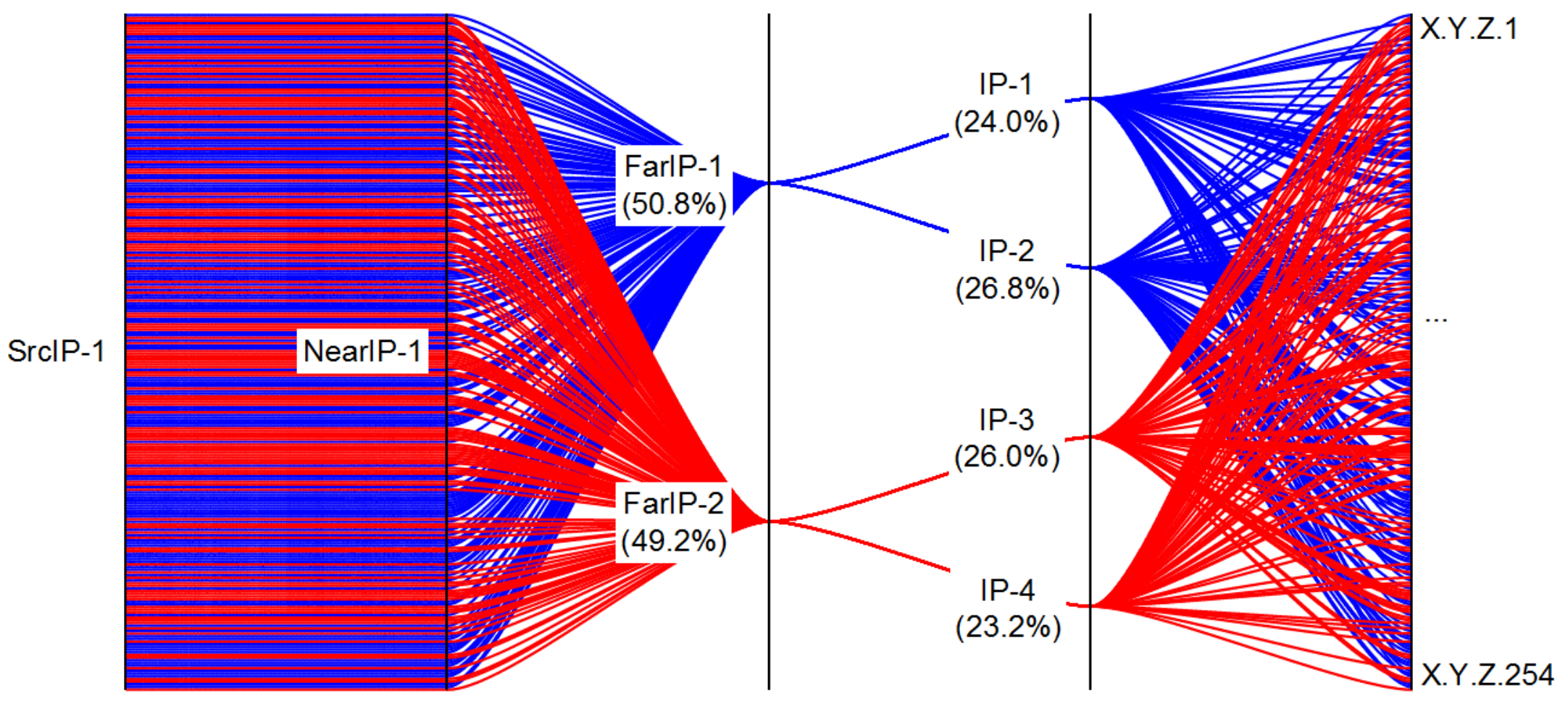}
     \label{fig:IXPCase-c}
     }
     \subfigure[Routing map from {\bf SrcIP-1} with ICMP  at 10:30GMTam]{
    \includegraphics[width = 0.47\textwidth] {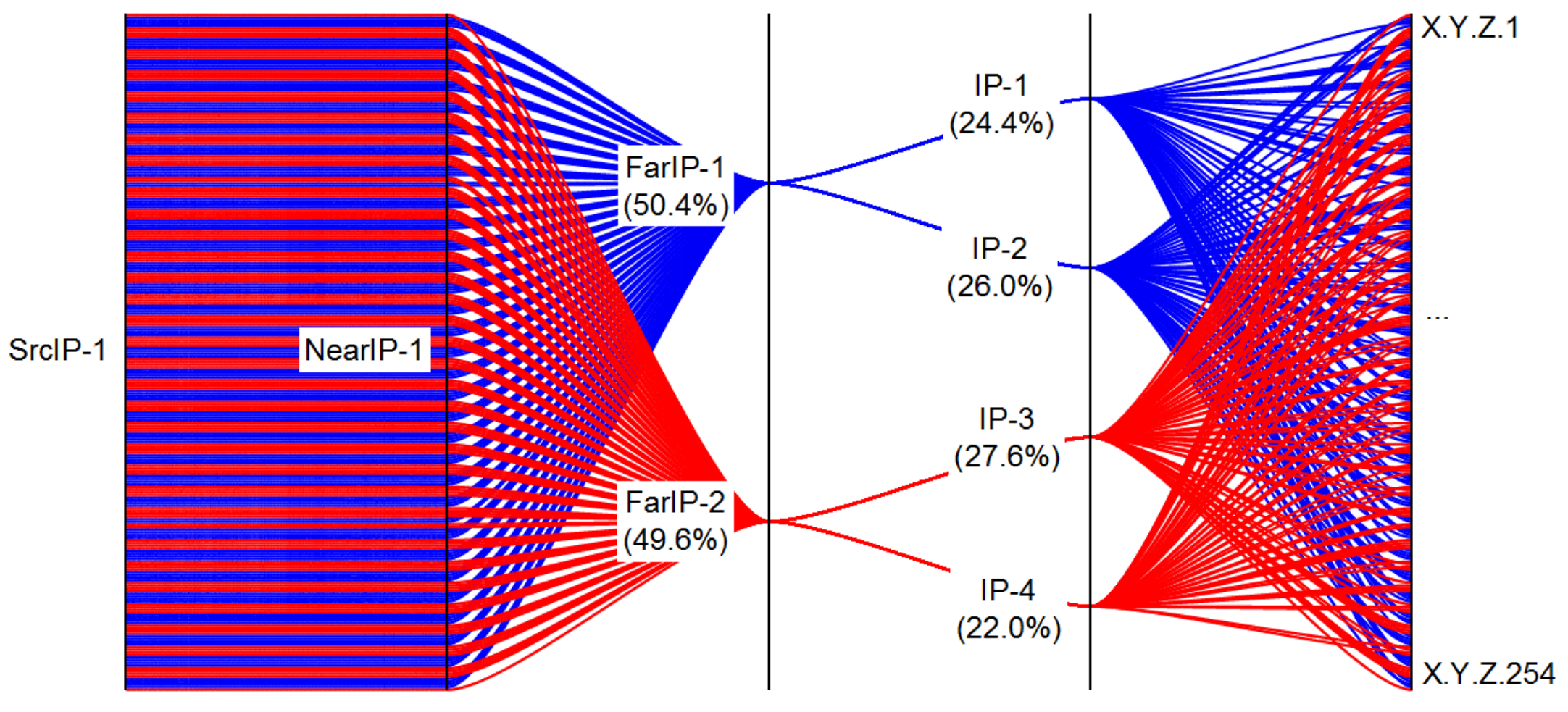}
     \label{fig:IXPCase-d}
     }
     \subfigure[Routing map from {\bf SrcIP-2} with ICMP at 10:30GMT]{
    \includegraphics[width = 0.47\textwidth] {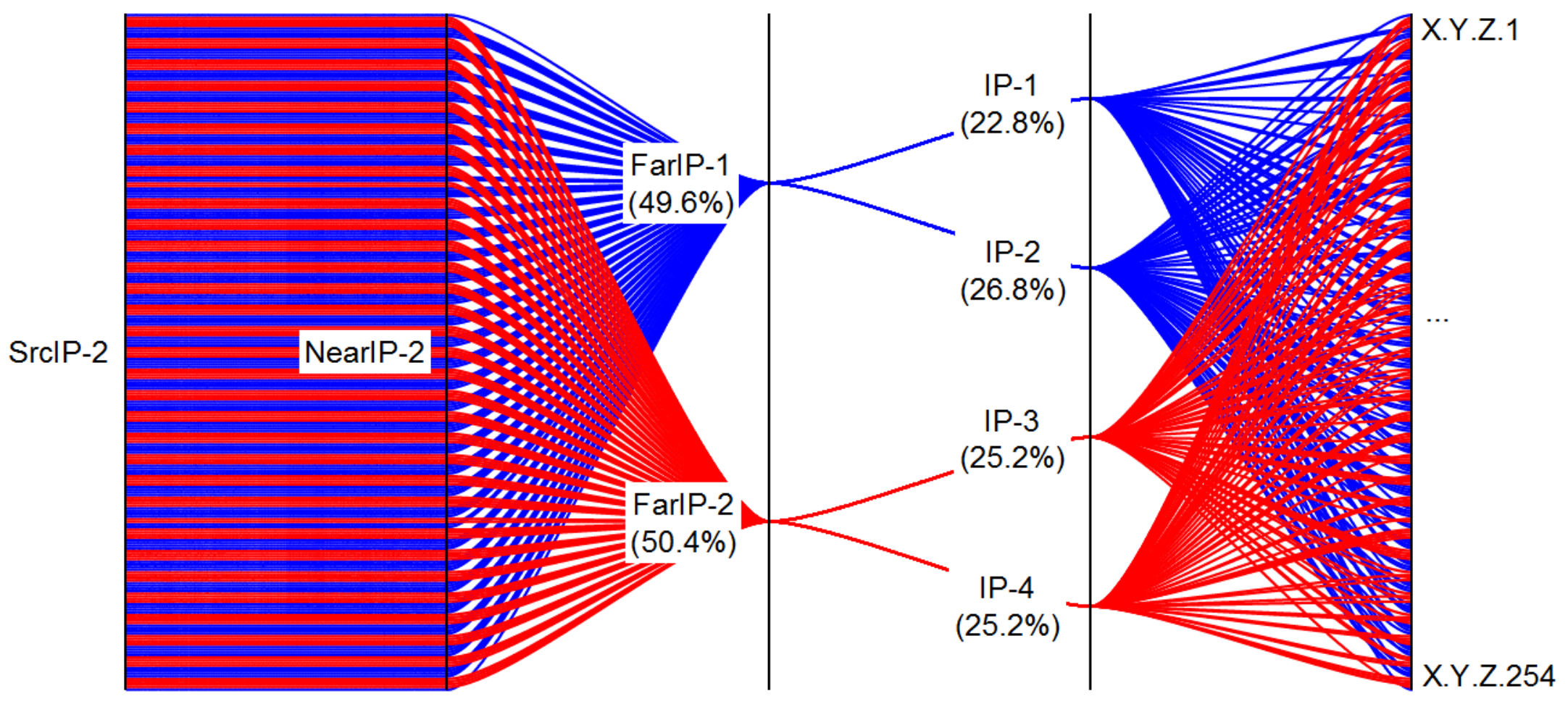}
     \label{fig:IXPCase-e}
     }
    \caption{Topology and routing maps for BGP-M case <AS6939, {\tt tyo1}, AS2907, 160.18.2.0/24>. 
    (a) topology map shows  connectivity of the BGP-M case. 
    (b) -- (e) routing maps illustrate traceroute paths to each IP address in the destination prefix.
   (b) and (c) were based on UDP packets from the same source (SrcIP-1 at 209.51.186.5)  but measured at different times in the morning of 20/May/2021. 
   (d)~and~(e) were based on ICMP packets measured at the same time but from two different sources (SrcIP-1   and SrcIP-2 at 65.19.151.10). 
   The two border links have the same bandwidth of 100GB~\cite{PeeringDB}.
     }
    \label{fig:IXPCase}
\end{figure*}

\subsection{Load Balancing Algorithms of Cisco Routers}

For all BGP-M cases that we were able to run traceroute measurement, the LG commands ({\tt routes} and   {\tt summary}) and the  returned information were all in the style and format of Cisco, indicating the {\em NearBRs} where these BGP-M cases were implemented were all Cisco routers.

According to Cisco's configuration documentation~\cite{CISCO-LB}, by default,  Cisco routers can configure BGP-M as per-session  load balancing,  where traffic allocation decisions are made by hash algorithm for each pair of source and destination IP addresses. 
Network operators can also configure their routers as per-packet round robin load balancing, or per-flow load balancing where the hash algorithm considers  source and   destination IP addresses and port numbers.

Figure \ref{fig:IXPCase-a} shows the topology map  of  the BGP-M case  <AS6939, {\tt tyo1}, AS2907, 160.18.2.0/24>,  where the {\em NearAS} connects with the {\em FarAS}  via two border links. 
We sent traceroute packets from two RIPE Atlas probes located inside AS6939, i.e.\,SrcIP-1 and SrcIP-2. 
Traffic from the two sources arrived at the {\em NearBR} ({\tt tyo1}) at two different ingress interfaces, i.e.\,{NearIP-1} and {NearIP-2}. 
Traffic from each source was shared on the two border links, i.e.\,BL-1 and BL-2. 
According to the IXP data from PeeringDB~\cite{PeeringDB}, FarIP-1 and FarIP-2  belong to  two IXPs, named as JPIX TOKYO and JPNAP Tokyo, respectively.

Figures \ref{fig:IXPCase-b}-\ref{fig:IXPCase-e} show the routing maps observed from traceroute probings with different settings.  
Figures \ref{fig:IXPCase-b} and \ref{fig:IXPCase-c} are both based on UDP packets sent from SrcIP-1 (209.51.186.5), but at two time points 15 minutes apart. 
Figures \ref{fig:IXPCase-d} and \ref{fig:IXPCase-e} are both based on ICMP packets at the same time point, but sent from different sources: SrcIP-1 (209.51.186.5) and SrcIP-2 (65.19.151.10), respectively. 
Here are some  observations. 

\subsubsection{Firstly} all the four routing maps show that probes to the IP addresses in the destination prefix were always equally shared on the two border links, which, as expected, showed BGP-M provides load balancing at the level of destination prefix.  %

\subsubsection{Secondly} on the routing maps in Figures \ref{fig:IXPCase-b} and \ref{fig:IXPCase-c} based on UDP packets, the packets to different destination IPs were randomly allocated on the two border links, and the allocations varied at different time points.
This is the hallmark of load balancing based on the so-called {\em include-ports}  algorithm~\cite{CISCO-LB}, which takes into account  IP addresses and port numbers of source and destination. 
While hash function is sensitive to any change of bits in the identifiers, the UDP packets have not only different destination IPs, but also different port numbers when sent at different time points.

\subsubsection{Thirdly}  on the routing maps in \ref{fig:IXPCase-d} and \ref{fig:IXPCase-e}, the ICMP packets were allocated on the two border links in a regular way: packets to  4 consecutive destination IPs were allocated  on one border link, and  the next 4 on the other border link; then the pattern repeated alternately. 
This suggests (1) the Cisco router was configured to conduct per-session load balancing for ICMP traffic using the so-called {\em universal} algorithm~\cite{CISCO-LB} which considers only  source and destination addresses; and (2)~only a part of the destination IP address was considered~\cite{Almeida2020INFOCOM}.
 
Closer inspection revealed that the BGP-M allocation patterns in the two routing maps were exactly opposite to each other, i.e.\,destination IPs allocated to BL-1 in \ref{fig:IXPCase-d} were allocated to BL-2 in \ref{fig:IXPCase-e}, and vice versa. 
This is because the routing maps were based on packets sent from different source addresses. 
Indeed, due to the {\em universal} algorithm, there are only two possible allocation patterns for ICMP packets  from any sources to IP addresses in a destination prefix.

Cisco routers implement load balancing for UDP and ICMP traffic in vastly different ways. 
Since most real traffic flows are TCP or UDP, we   should conduct traceroute measurements with UDP packets to reveal the true picture of BGP-M load balancing.

\subsection{Diverse Routing Patterns in {\em FarAS}}

The topology map in Figure \ref{fig:IXPCase-a} shows that in this BGP-M case, the two border links entered the {\em FarAS} via two different IXPs, each of which further split traffic onto two internal links within the {\em FarAS}. 
The routing maps in Figure \ref{fig:IXPCase} show that   the two IXPs conducted load balancing in a random way so that the internal links  received similar portions of traffic, whether probed by UDP or ICMP from the same or different sources.

Due to limited availability of RIPE Atlas probes located in relevant nearside ASes,  we were only able to run traceroute measurement\footnote{With the measurement standard we set in Section VI-A.} with ICMP packets for 89 of all BGP-M cases uncovered in this study.

{We observed the followings: (1) in 33 cases, traffic on each border link was further split onto different links in the {\em FarAS}; (2) in 22 cases,   traffic on different border links were forwarded separately to destination prefix via parallel routing paths; (3)  in 14 cases, traffic on all border links were later merged into a single path;   (4)  in  15 cases,    there were  complex routing in  the {\em FarAS}, and (5) in the other 5 cases, traceroute contained unresponsive hops. 
We  conducted traceroute with UDP packets and observed similar results.  }

The above observation on diverse intra-domain routing in {\em FarAS} suggests that the deployment of BGP-M at border routers of an AS is independent, or transparent, to  intra-domain load balancing in other ASes.

\begin{figure*}[]
    \centering
      \subfigure[{\bf ICMP:} Distribution of the link delays.    ]{
    \includegraphics[width = 0.47\textwidth] {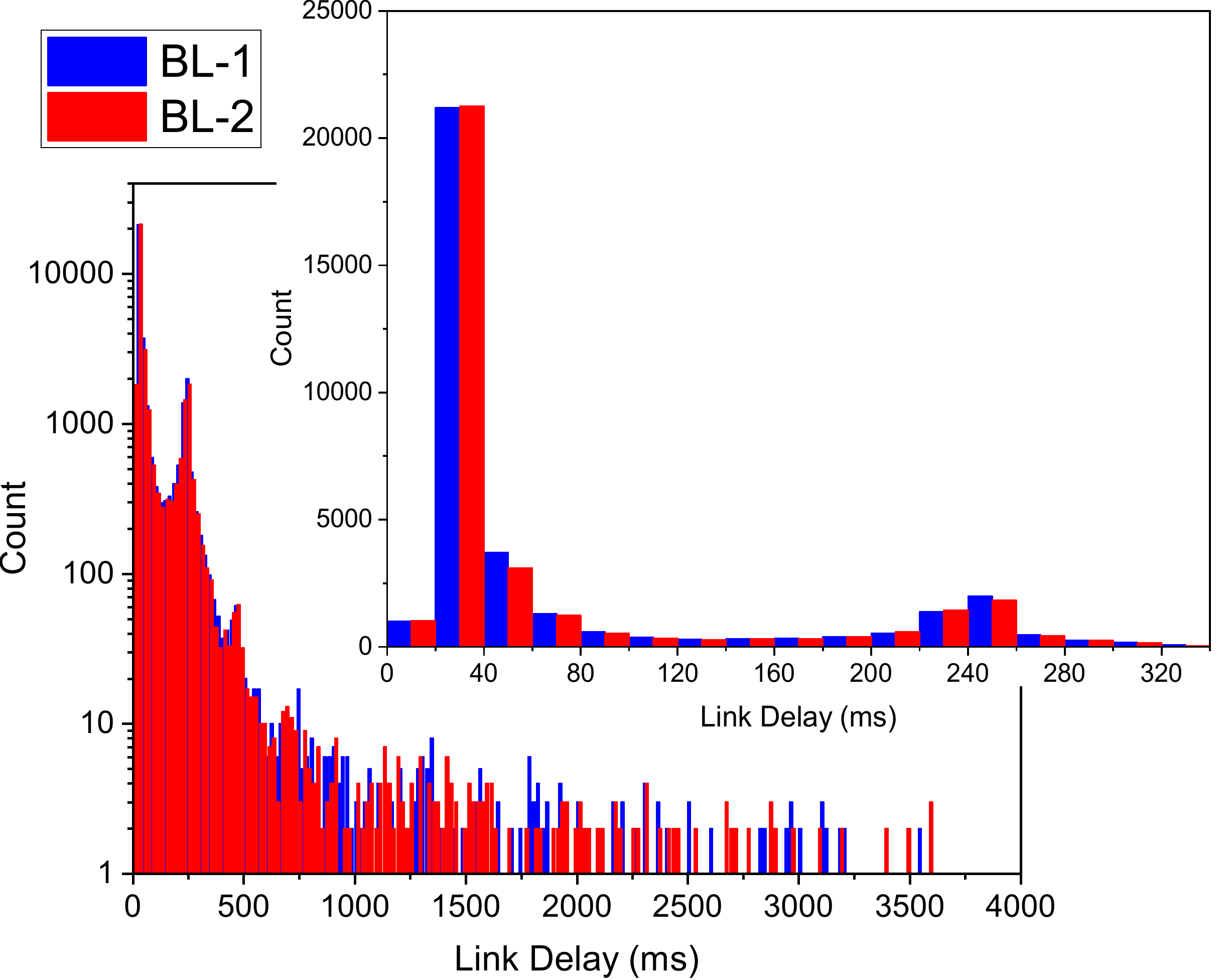}
    \label{fig:caseDelays-a}}
     \subfigure[{\bf UDP:} Distribution of the link delays.]{
    \includegraphics[width = 0.47\textwidth] {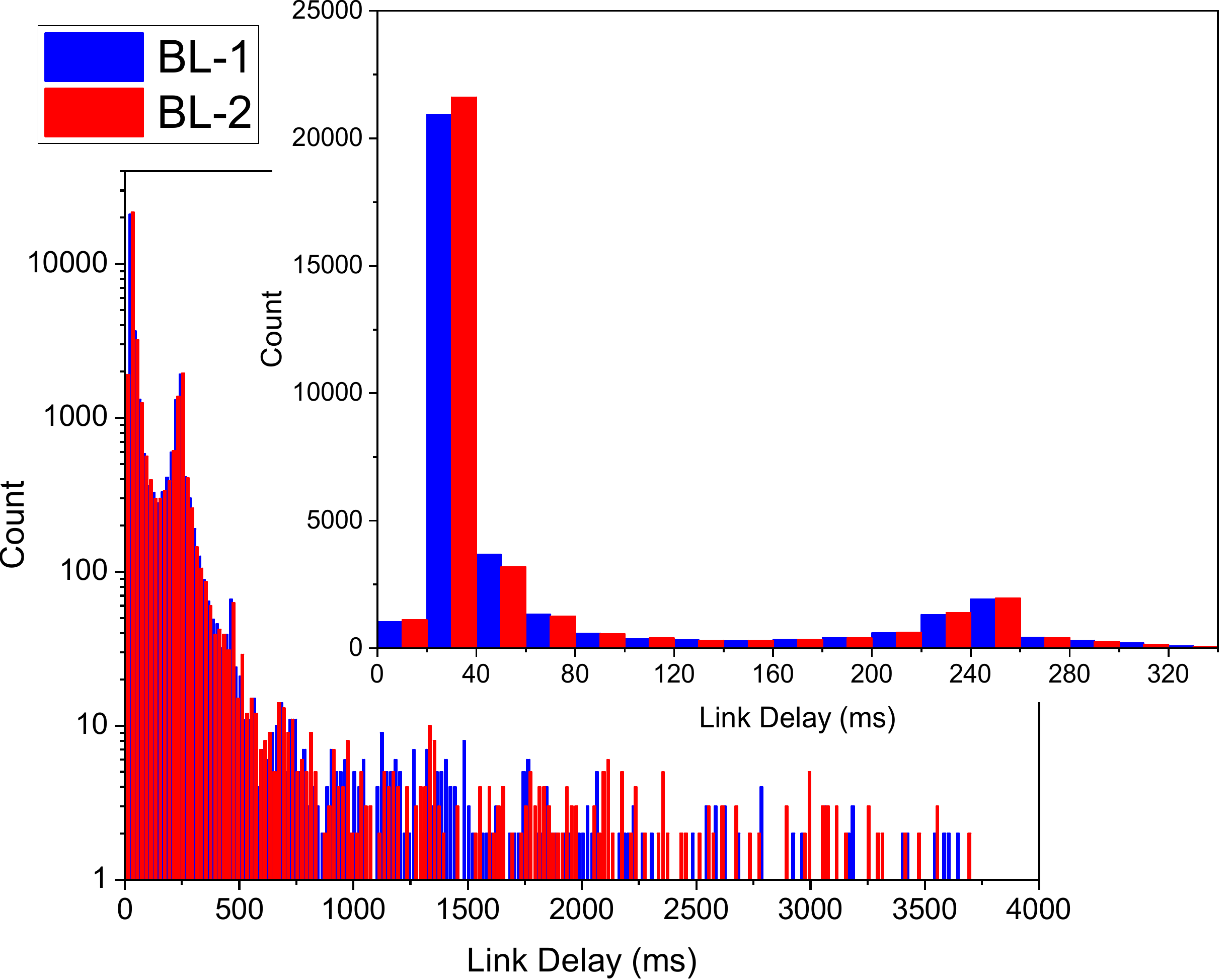}
    \label{fig:caseDelays-b}}   
    
    \subfigure[{\bf ICMP:} Link delay at 15-minute intervals over 3 days.]{
    \includegraphics[width = 0.47\textwidth] {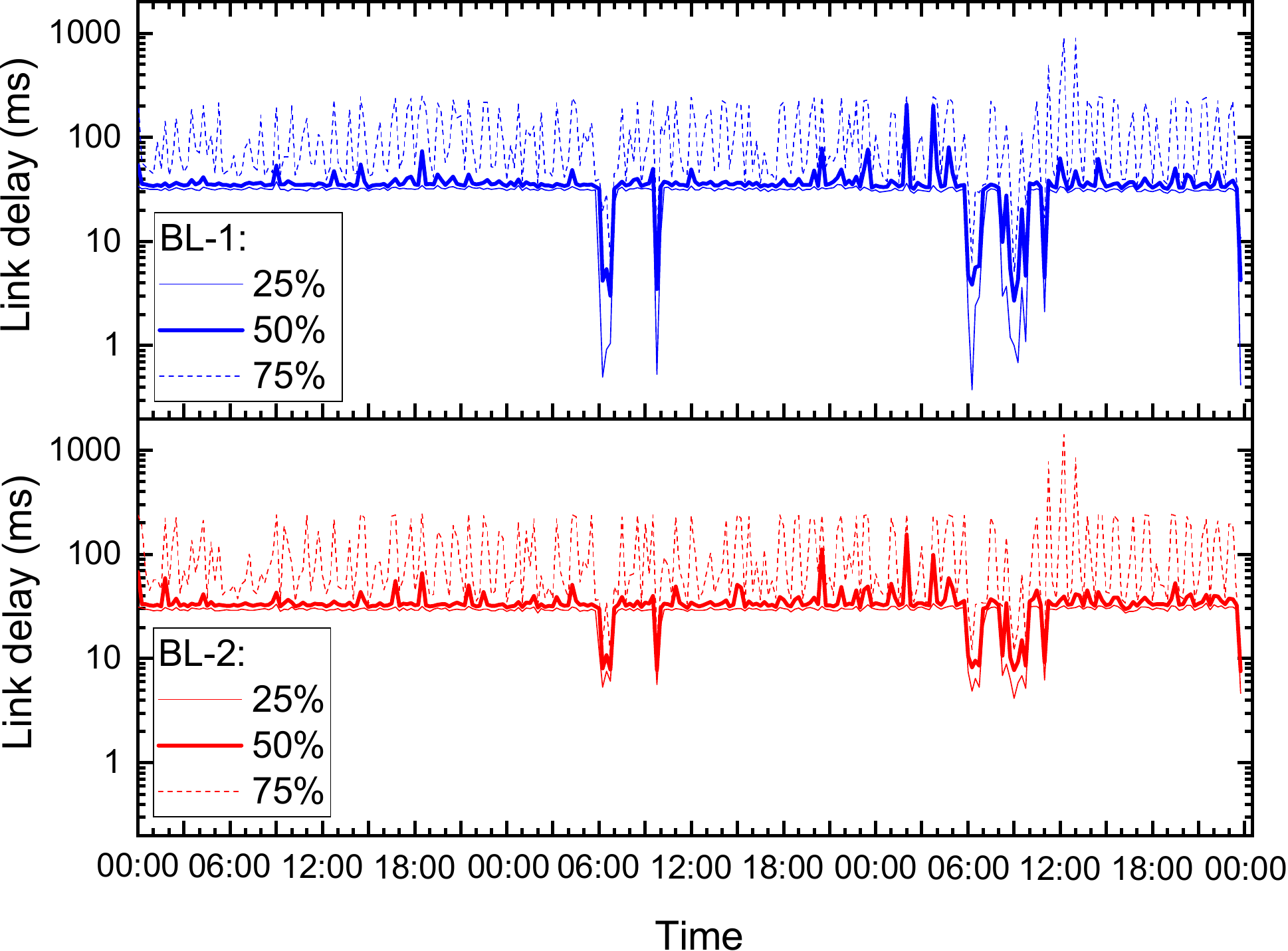}
    \label{fig:caseDelays-c}}
     \subfigure[{\bf UDP:} Link delay at 15-minute intervals over 3 days.]{
    \includegraphics[width = 0.47\textwidth] {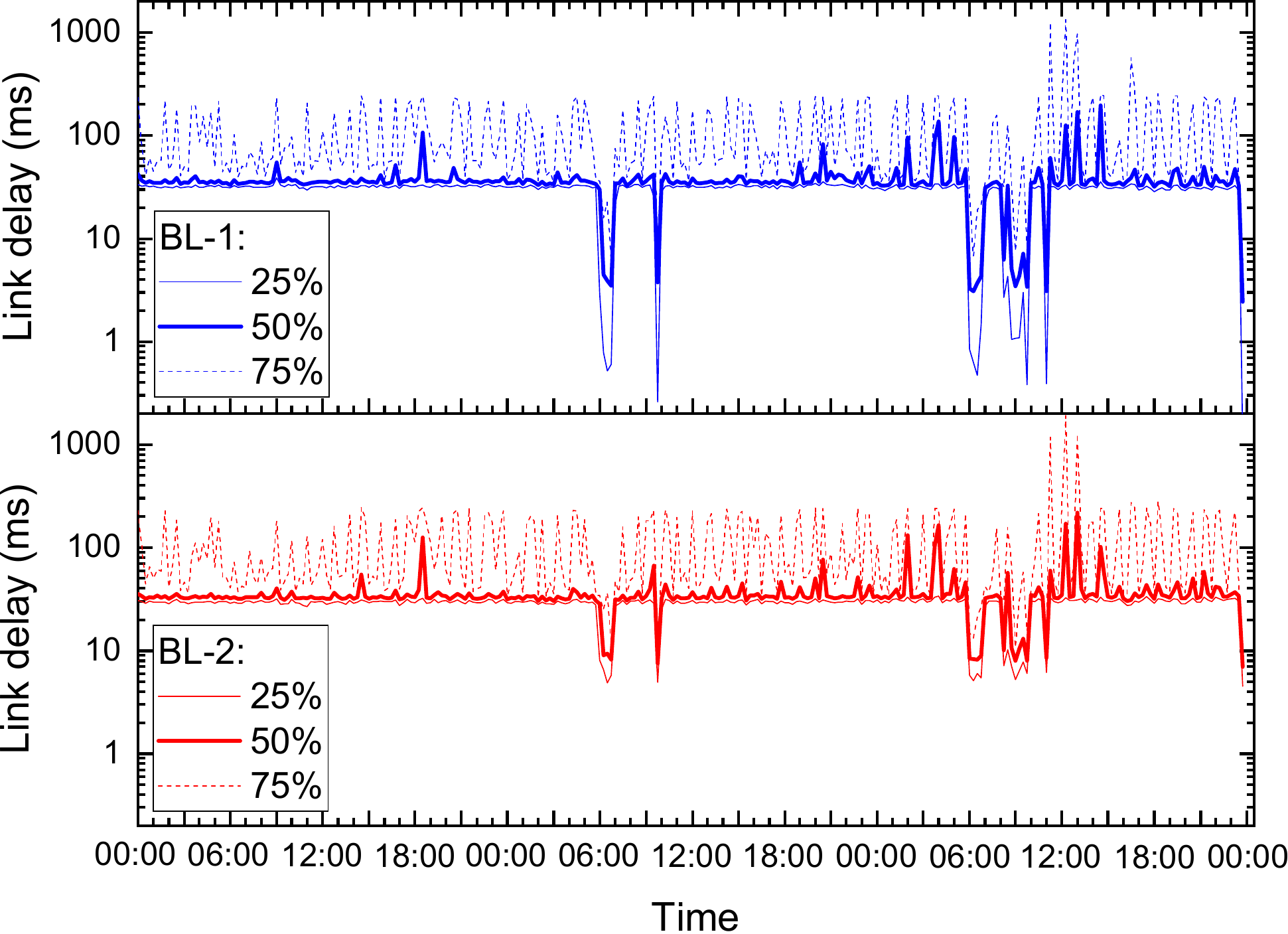}
    \label{fig:caseDelays-d}}
    \subfigure[{\bf ICMP:} Link delay   for each destination IP]{
    \includegraphics[width = 0.47\textwidth] {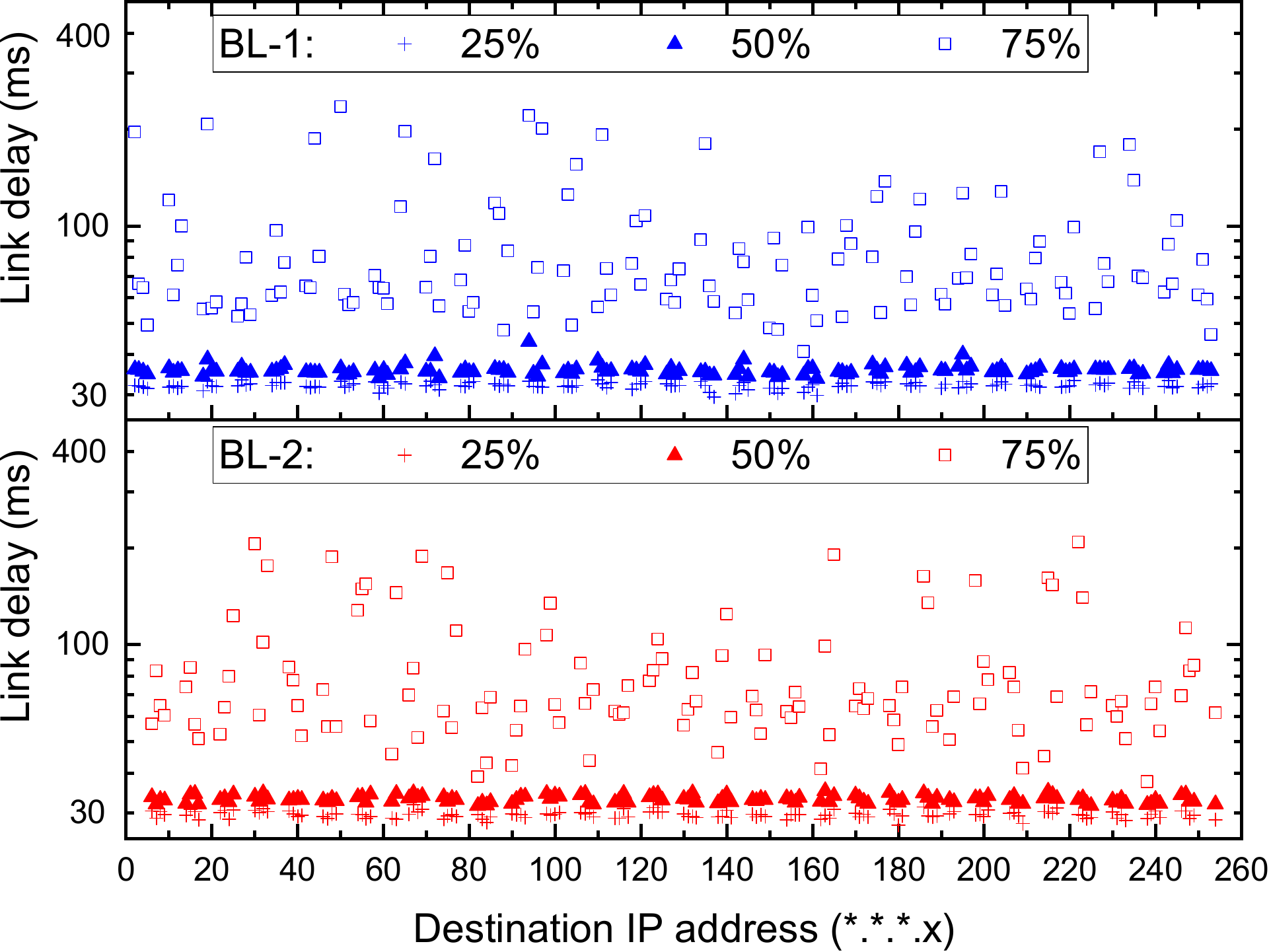}
    \label{fig:caseDelays-e}}
    \subfigure[{\bf UDP:} Link delay   for each destination IP]{
    \includegraphics[width = 0.47\textwidth] {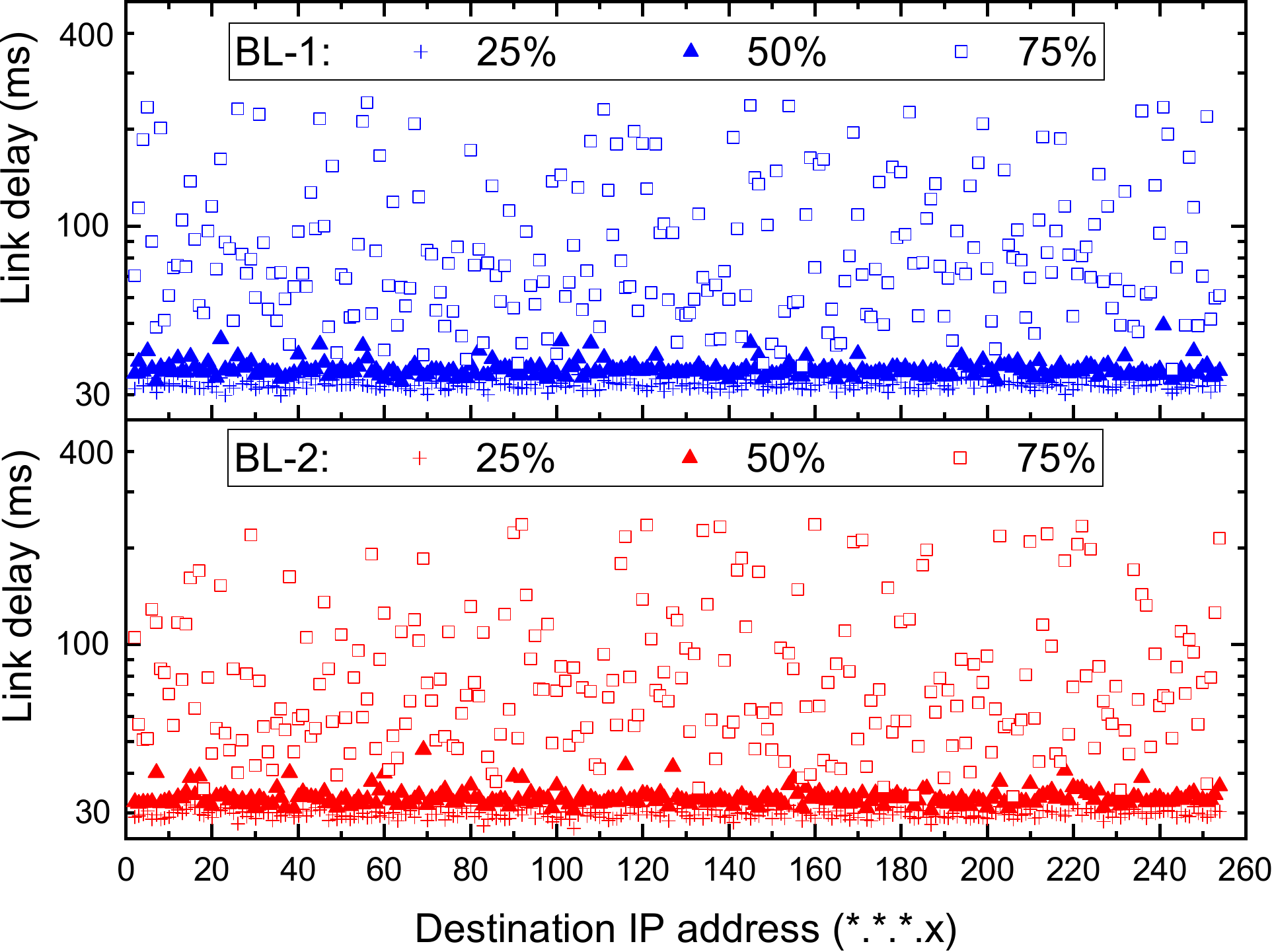}
    \label{fig:caseDelays-f}}

    \caption{Delays  on the two border links of BGP-M case <AS6939, {\tt hkg1}, AS20940, 23.67.36.0/24>.  Delays were measured by sending traceroute   ICMP and UDP packets (with one minute separation) from a RIPE Atlas probe in AS6939 to each IP address in the destination prefix 23.67.36.0/24  at 15-minute intervals starting from 00:00am GMT on 16 June 2021 for 3 days.  Y-axis is plotted on log scale. 
    {\bf (a) and (b)} show the distribution of link delay  values (in 20ms bins)  to all IP destination   at all time points,  where the inset shows the distribution (in linear scale) of   delays between 0ms and 340ms that account for >95\% of all values. 
    {\bf (c)~and~(d)} show   statistics of link delay   calculated over link delay values to all  destination IPs at a given time point; 
    {\bf (e)~and~(f)} show statistics calculated over values to a given destination IP at all time points.
    }
    \label{fig:caseDelays}
\end{figure*}

\section{Measurement of Delays on Border Links }
\label{Sec:Delays}

To understand the effectiveness and performance of load balancing by BGP-M, here we study traffic delay on BGP-M border links based on traceroute measurements using ICMP and UDP packets\footnote{Our recent work reported in~\cite{Li2021GI} used only ICMP packets.}. 

\subsection{Measuring Link Delay Based on Traceroute RTT}

Figure \ref{fig:caseDelays} shows the link delays measured on the two {\em{BLs}} for the BGP-M case  <AS6939, {\tt hkg1}, AS20940, 23.67.36.0/24>, which was deployed by HE (AS6939) at its border router {\tt core1.hkg1.he.net} (with {\em{NearIP}} 184.105.64.129) to the {\em{DstPrfx}}  of 23.67.36.0/24 in a neighbour AS called Akamai (AS20940).  
We sent traceroute using ICMP packets and UDP packets, with one-minute separation, from a RIPE Atlas probe to each of 254 IP addresses in the destination prefix at 15-minute intervals for 3 days from {00:00am Hong Kong local time (i.e.\,GMT + 08:00) on 16 June 2021}.
We used default RIPE Atlas traceroute settings.

From each traceroute measurement to a destination IP,  we obtained the Round Trip Time (RTT) value at each IP hop; and then we calculated the {\em delay} on a border link, which was the difference between the RTT values of {\em NearIP} and {\em FarIP} of the border link.
 
The link delay  includes the following sources:   (1)~processing delays at   {\em NearBR} and {\em FarBR}, which are negligible because of   border routers' high performance; (2)~serialisation delay at {\em NearBR}, which is  negligible because of   border links' high bandwidths; (3)~transmission delay  on  border links, which is negligible because of   small distance (<30km) between the relevant facilities, all located in Hong Kong; and (4)~queuing delay at {\em NearBR}, which accounts for most of the link delay. 
Since the link delay mainly measures the queuing delay at the {\em NearBR}, it reflects  the level of  traffic congestion  for each of  the border links  and therefore can be considered as an indicator of routing performance. 

As explained in the previous section,  when using ICMP packets, traceroute paths to a given destination IP at different time points always go through the same border link due to per-session load balancing that considers only source and destination addresses; whereas for UDP packets, traceroute paths to a given destination IP at different time points are allocated to any of the two border links at random due to per-flow load balancing that considers IP addresses and port numbers of source and destination.

\subsection{Congestion-free Transit on Border Links}

Figures \ref{fig:caseDelays-a} and \ref{fig:caseDelays-b}  show the distributions of link delays on the two border links to the destination IPs measured by 15-minute interval during three days based on ICMP and UDP packets, respectively. 
In general, the link delay values were mostly small, mainly  between 20ms and 40ms. 
Although there were  delays more than a few seconds, they were very rare.
This indicates that transit on these border links were mostly free of congestion.
 
The congestion-free transit can be explained by the large bandwidths and relatively small traffic volumes on the two border links.
The bandwidths of the two border links were: 10G for BL-1 (with FarIP-1 103.247.139.17) and 100G for BL-2 (with FarIP-2 123.255.91.169). 
We obtained the bandwidth information from PeeringDB~\cite{PeeringDB}, where the public peering data for Akamai (AS20940) was last updated on 10/July/2021.
We also obtained from Akamai's technical  report\footnote{\url{https://www.akamai.com/us/en/multimedia/documents/state-of-the-internet/q1-2017-state-of-the-internet-connectivity-report.pdf}} that Akamai at Hong Kong (where the {\em FarBRs} were located) had  average traffic volume of 21.9\,Mbps and peak volume of 129.5\,Mbps in Q1 2017. 
Although the report was four years ago, today's traffic volume is likely to remain well below the bandwidths of the border links.  

\subsection{Load Balancing on the Border Links}

Figures \ref{fig:caseDelays-c} and \ref{fig:caseDelays-d} show the change of link delay  on  the two border links at different time points in 3 days. 
For each time point, we show the median as well as the   25th and 75th percentiles of  calculated link delay values to all destination IPs. 

Figures \ref{fig:caseDelays-e} and \ref{fig:caseDelays-f} show the statistics of link delay  to each IP address in the destination prefix. 
As explained in previous section,  ICMP packets to destination IPs are equally allocated to the two border links in exactly the same way at every time point. 
That is, in Figure \ref{fig:caseDelays}(e), the statistics for BL-1  show link delays to  only 128 destination IPs, each of which was calculated from 288 measurements (= 3\,days $\times$ 24\,hours $\times$ 4\,times/h); whereas the statistics for BL-2 show link delays to 125 different destination IPs\footnote{No traceroute data to 23.67.36.1.}. 
By comparison, UDP packets to destination IPs are equally, but randomly, allocated to the two border links, and  allocation changes randomly at every time point.
Thus, in Figure \ref{fig:caseDelays}(f), the statistics for both BL-1 and BL-2  show link delays to  all of 253 destination IPs, each of which was calculated from  measurements at about half of the time points. 

The two border links always had similar delay values, throughout the duration of measurement,  to all destination IPs, and for  both ICMP and UDP packets. 
This is as expected. 
It vividly illustrates the desired result of BGP-M load balancing, where traffic is equally   shared between   border links to fully utilise  routing capacity and diversity available at borders of ASes, with the purpose to reduce congestion and improve routing flexibility and resilience.

\section{Extraordinary Effort of HE in BGP-M}
\label{Sec:Effort}

According to the CAIDA AS Rank data~\cite{caida_asrank},  Hurricane Electric (HE, AS6939) is the 7th largest AS in terms of customer cone size, and it provides transit between more than 8k ASes, which account for 12\% of all ASes in global routing tables. 
HE is also a hyper-giant AS of very high peering affinity and port capacity~\cite{hyper-giants}, with more than 6k peers and presence at 236 IXPs -- more than all other ASes. 
Hence, HE has a particular need for the best practice of traffic engineering in order to achieve stable, reliable and high routing performance with its enormous number of neighbours. 

\subsection{Extensive Deployment of BGP-M Cases}

Our results show that HE has  extensively deployed BGP-M across its entire network. 
It has deployed BGP-M to 611 (i.e.\,>10\%) of its neighbour ASes at 69 (i.e.\,>60\%) of its border routers on the IPv4 Internet. 
HE has also widely deployed BGP-M on the IPv6 Internet. 
Notably, HE deployed large numbers of BGP-M cases  to other hyper-giant ASes, especially   large content providers, on both IPv4 and IPv6 Internet (see Section \ref{Sec:BGP-MDeployment_HE_NeighbourASes}).
We  have confirmed with HE on their wide deployment of BGP-M   through our consultation with their network operators.

The fact that a large, traffic-intensive, transit network like HE has devoted such   extraordinary  effort in wide deployment of BGP-M across its world-wide network is an evident  indication of the significant benefit and advantage that this load balancing  technique can provide. 

\subsection{Active Maintenance of BGP-M Cases}

As shown in Figure \ref{fig:AS6939BGPRoute}, LG server's  response   to   the {\tt routes} command   ({\tt show ip bgp routes detail <IP address>})  contains rich details on {the deployment of BGP-M}, including the  time since the routing table has been last updated. 
This allows us to track  the changes of a BGP-M case   by re-querying the relevant border router at a later time. 

As shown in Table \ref{tab:Reasons}, at the beginning of July 2021, we revisited each of BGP-M cases of HE that we observed in our 2020 measurement. 
 
For the 1,088 cases deployed by HE on the IPv4 Internet, 632  (or 58\%) of the cases remained exactly the same; and 60 cases  had additional or replaced border links, which, according to our definition, were still of the same BGP-M cases as they were deployed at the same border routers for the same destinations. 
We also observed that 396 (or 36\%) cases were either disappeared or changed since our 2020 measurement. 
For example, LG queries suggested some nearside border routers were `Not existing' anymore, and some routes were not labelled as `M' (i.e.\,multipath) anymore. 
A small number of cases were still there but with changed attributes making them different or new BGP-M cases, for example with a changed farside AS. 
We observed similar results for the BGP-M cases on IPv6.

The above observations suggest   that HE has been actively maintaining  and rearranging its BGP-M cases.
Some of the changes might occur in reaction to network changes while others were likely for the purpose of achieving better configuration to gain more benefit from BGP-M load balancing. 

\begin{table}[]
    \centering
    \caption{Changes of BGP-M cases deployed by HE on   IPv4 and IPv6 Internet}
    \begin{tabular}{l|c|c}
    \hline
                      &    IPv4   &   IPv6 \\ \hline\hline
         2020 measurement dates & Jan-May 2020 & July-Oct 2020\\  
         Total \# of BGP-M cases & 1,088 & 300 \\ \hline\hline
         2021 measurement date & July 2021 & July 2021 \\ \hline
                & \multicolumn{2}{c}{\# of  remaining cases}   \\  \hline
                 Total  \# of  cases & 692  & 218 \\ 
               Exactly   same as before & 632  & 204 \\
        Increased \# of BLs   & 33    & 7  \\
        Same \# but different BLs  & 27    & 7  \\ \hline
                & \multicolumn{2}{c}{\# of  disappeared/changed  cases}   \\  \hline
        Total  \# of  cases                     & 396   & 82  \\
        {\em NearBR} `Not existing'  & 13    & 0  \\
        `No routes' for {\em DstPrfx} & 143   & 12 \\
        Status without `M' (multipath)                 & 109   & 25 \\
        Status without `E' (eBGP)                & 102   & 32  \\
       Other changes        & 29    & 13 \\  
        \hline
    \end{tabular}
   
    \label{tab:Reasons}
\end{table}

\section{Advantages and Benefits of BGP-M  }
\label{Sec:Benefits}

As a load balancing technique, BGP-Multipath not only provides balanced traffic and enhanced routing performance, but also offers a number of unique advantages and benefits in terms of deployment and operation. 

\subsection{Wide Availability and Readiness for Implementation}
\label{Sec:Benefits_Availability}
Both hardware and software requirements  for {the deployment of BGP-M} are already widely available  in the Internet. 

Firstly, there is a wide presence of  multiple border links between ASes in the Internet, where more than one border links are connecting from a  border router of an AS to border router(s) of a neighbour AS. 
Such multiple border links   commonly exist, especially among core ASes or between core and peripheral  ASes. 

Secondly, most border routers provided by major router vendors, such as Cisco, Juniper and Huawei,  already support BGP-M load balancing, which is an integral part of their design and function. 

This means BGP-M can be readily implemented by   network operators with many of their neighbours without    changing or upgrading their infrastructure or agreements.  

\subsection{Easy  Implementation}
\label{Sec:Benefits_Implementation}
The implementation of BGP-M is rather simple and straightforward. For example, the minimum action required on a Cisco border router is to activate   BGP-M by changing a single parameter {\tt maximum-paths} from its default value 1 to the number of (different) paths for a given {\em DstPrfx}. 
There is literally no additional cost to implement BGP-M.

\subsection{Independent, Flexible and Transparent Deployment}
\label{Sec:Benefits_Deployment}
Although we call it BGP-M and the technique follows BGP's best path selection process, network operators do not need to alter their BGP process to deploy BGP-M as the load balancing will still follow exactly the same AS-level path as before. 
As such, network operators can freely and independently implement or remove BGP-M without informing or obtaining new agreement from their neighbour ASes. 
Network operators can deploy, revise and cancel BGP-M for any selections of destination prefixes in any neighbour ASes. 

{There is no interference between {the deployment of BGP-M}, any other multipath routing techniques implemented  within or outside of the AS, and any  traffic engineering configurations elsewhere.} 
For example, as we showed in previous sections, there is no impact on BGP-M load balancing whether the border links connect to the neighbour AS directly or via IXPs, or whether and how IXPs further apply their own load balancing arrangement. 

Basically, BGP-M deployed at a border router is transparent to other parties participating in the relevant traffic routing, which gives network operators flexibility and convenience.

\subsection{Benefits of BGP-M Load Balancing}
\label{Sec:Benefits_Benefits}
The benefits of load balancing gained from {the deployment of BGP-M} is no less than any other multipath routing techniques. 
It can increase more balanced use of border links and reduce risk of congestion in face of traffic surges. 
It can also improve routing path diversity, which can be useful   for network resilience and security. 

{In addition, the border links are common, already there, and shared by networks.} 
Many border links  have high bandwidths. 
It makes sense to fully utilise these resources that are readily available, especially when it is easy and convenient to do so.  
A network operator benefits from  {the deployment of BGP-M} regardless of whether or how many other networks have implemented the technique. 
The more deployment, the more benefit. 
And such benefits are likely to be mutually beneficial to not only the AS that deploys BGP-M but also its neighbour ASes. 

\subsection{Immense Potential for Future {Deployment of BGP-M}}
\label{Sec:Benefits_Potential}
Although our work shows  BGP-M has already been widely deployed, as explained in previous sections, our inference is a conservative, lower bound estimate of the scale of {the deployment of BGP-M} in the Internet and there could be many more BGP-M cases. 
Indeed we  have recently discovered a few hundreds more BGP-M cases by querying LG servers all  prefixes {(of all sizes)} announced by  each neighbour AS. 
Many of them are BGP-M cases implemented by HE to other hyper-giant ASes.
Considering only a relatively small portion of ASes provide LG services, there should be   more ASes that have implemented BGP-M in the Internet. 

Nevertheless, based on our data and analysis so far, we estimate that the  scale of  existing {deployment of BGP-M} is still far smaller than the intra-domain multipath  routing, of which millions of cases  have been uncovered throughout the Internet. 
This means there is an immense scope for future deployment of BGP-M by  more ASes to  more destinations. 

In recent years, a significant amount of  investment and  effort have been devoted in coping with the rapid increase of traffic volume in the Internet.  
This study provides a technical and economic case for more deployment of BGP-M as an  option for load balancing, which is compatible and complementary to other traffic engineering techniques.

\section{Conclusion and Future Works} 
\label{Sec:Conclusion}

\subsection{Conclusion}
\label{Sec:Conclusion_Contributions}
This paper reports the first measurement of {the deployment of BGP-M} in the Internet. 
Our measurement was based on Looking Glass server data, which provides not only the ground truth, but also rich information on various aspects of BGP-M. 

We   provided the state-of-the-art knowledge on the BGP-M. 
We   ran traceroute from RIPE Atlas probes to reveal the exact   patterns of traffic allocation  as well as delays on border links. 
We provided in-depth analysis on a series of BGP-M cases deployed by Hurricane Electric, as a capital example of large-scale deployment of BGP-M by a major transit network. 

Our work is valuable to   network operators 
interested in load balancing. 
{ 
For example, our work and the example of HE may inspire more network operators to consider   deploying BGP-M and steering their traffic with  neighbour ASes via multiple border links.

Our work is also valuable to Internet researchers, who can use our measurement data to obtain a fuller and more detailed picture of the inter-domain routing between ASes. This can help the study on inter-domain congestion~\cite{Dhamdhere2018SIGCOMM}  and the study on Internet traffic map~\cite{Koch2021HotNets}.
}

We have shared on GitHub \cite{GitHub} the LG data and the traceroute datasets for all discovered BGP-M cases reported in this paper. 


{
\subsection{Future Works}
\label{Sec:Conclusion_Discussions}

As an ongoing research, our work can be improved and extended in several ways. 

Firstly, we plan to discover more BGP-M cases. One method is to study more ASes with LG servers, which requires access to more LG servers without invoking ASes' policies. Another approach is to conduct large-scale traceroute measurement similar to   recent studies on  multipath routing~\cite{Vermeulen2020NSDI, Almeida2020INFOCOM}, where the   challenge   is to achieve accurate AS border mapping.
The aim is to provide a more complete and accurate picture of the  deployment of BGP-M on the global Internet. It would be interesting to find out whether BGP-M has been more extensively deployed than we present in this paper. 

Secondly, with more  measurement data, we will study  the reasons and motivations for network operators to deploy, or not deploy, BGP-M when multiple border links are available. For example, we will investigate whether the deployment of BGP-M is related to traffic volume and AS relationships. Such study  may  provide new knowledge on BGP-M including its utility as well as  drawbacks and limitations.

Thirdly, we will further analyse the routing performance of BGP-M using comparative measurements. For example, we will assess performance of two border routers with the same topological connectivity (in terms of border links with neighbour ASes) but only one of them has deployed BGP-M.    We will also  monitor   traffic performance of a border router before and after it activates BGP-M, which  requires collaboration from   network operators. Such study will  enrich our understanding of BGP-M.  
}

\begin{figure}{\includegraphics[width=2in,height=2.5in,clip,keepaspectratio]{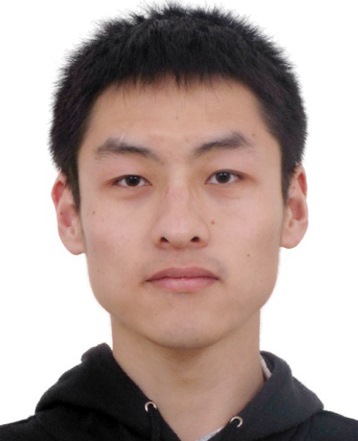}}
{Jie Li} received the B.S. and M.S. degrees from Dalian University of Technology, China, in 2011 and 2014, respectively. He received Ph.D. degree from University College London in 2021, where this research was conducted. 
His research interests include network measurement, BGP routing, mobile social networks, socially aware networking, and ad hoc networks. He is  currently  a senior engineer at Huawei, China with a focus on data centre network. 
\end{figure}

\vspace{1cm}

\begin{figure}{\includegraphics[width=3in,height=3.75in,clip,keepaspectratio]{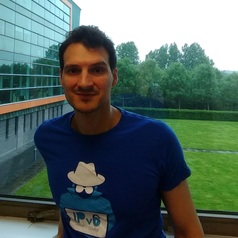}} {Vasileios Giotsas}
is a lecturer at Lancaster University, where he leads the Networks area research of the Security Institute.  He received his PhD from University College London (UCL). His research focuses on network measurements and the analysis of the routing system.
\end{figure}

\newpage

\begin{figure}{\includegraphics[width=2in,height=2.5in,clip,keepaspectratio]{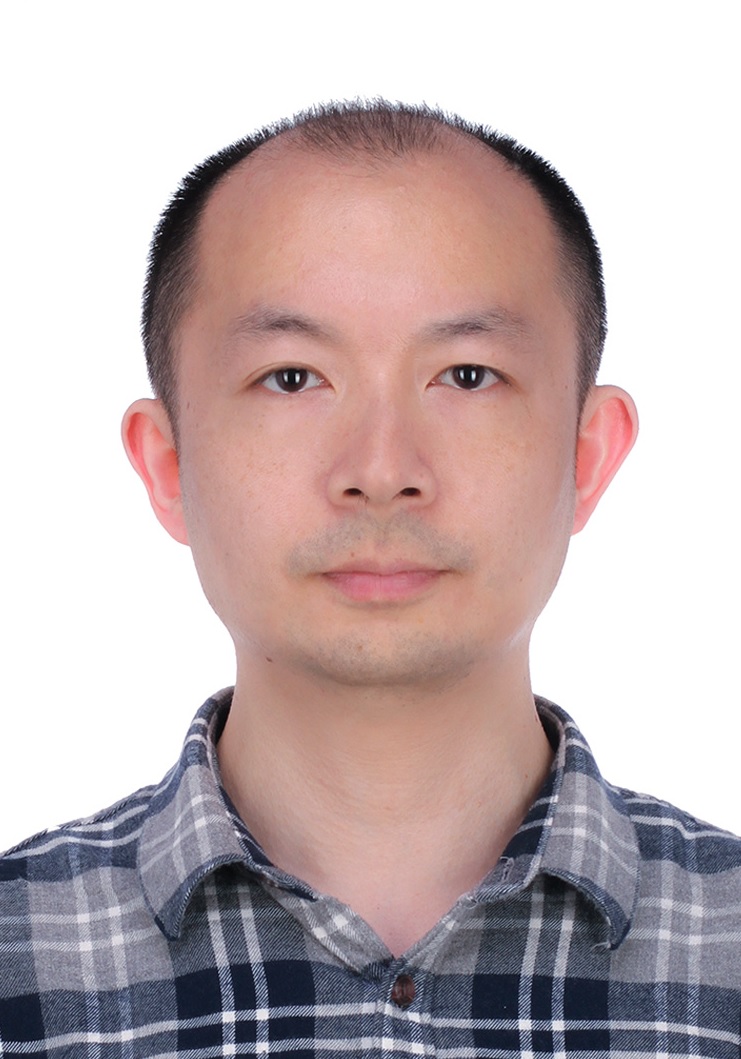}} {Yangyang Wang}
received the B.S. degree in computer science and technology from Shandong University, China, in 2002, the M.S. degree from Capital
Normal University, China, in 2005, and the Ph.D. degree from the Department of Computer Science, Tsinghua University, China, in 2013. He is currently
a research assistant professor, Institute for Network Sciences and Cyberspace, Tsinghua University. His current research interests include Internet routing architecture,  SDN/NFV, and network measurement. 
\end{figure}

\vspace{1cm}

\begin{figure}{\includegraphics[width=2in,height=2.5in,clip,keepaspectratio]{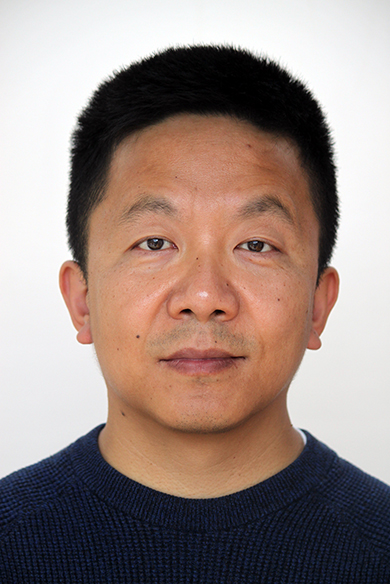}} {Shi Zhou}
(Senior Member, IEEE) received the B.Sc. and M.Sc. degrees in Electrical and Electronics Engineering from the Zhejiang University, China, and the Ph.D. degree in Telecommunications from the University of London, UK (QMUL) in 2004. He is an Associate Professor with the Department of Computer Science, University College London (UCL), London, UK. His research interests span the Internet routing and topology, complex networks, cybersecurity, cloud computing and online social media analysis. 
\end{figure}


\begin{thebibliography}{1}


\bibitem{RFC4271}
Y. Rekhter, T. Li, and S. Hares, ``A border gateway protocol 4 (BGP-4),'' RFC 4271, January 2006.


\bibitem{juniper-mbgp}
Juniper Networks, ``Understanding BGP-Multipath''. 
\url{https://www.juniper.net/documentation/en\_US/junos/topics/topic-map/bgp-multipath.html}

\bibitem{CISCO}
Cisco, ``BGP Best Path Selection Algorithm''.
\url{https://www.cisco.com/c/en/us/support/docs/ip/border-gateway-protocol-bgp/13753-25.html\#anc5}


\bibitem{huawei-mbgp}
Huawei, ``Example for configuring BGP load balancing''.  \url{https://support.huawei.com/enterprise/en/doc/EDOC1000178324/fd6029a9/example-for-configuring-bgp-load-balancing}

\bibitem{Vermeulen2018IMC} 
K. Vermeulen, D. S. Stephen , O. Fourmaux, and T. Friedman, ``Multilevel MDA-Lite Paris traceroute,'' in Proc. ACM IMC'18, pp. 29–-42.

\bibitem{Augustin2011TON} 
B. Augustin, T. Friedman, and R. Teixeira,  ``Measuring multipath routing in the Internet,'' IEEE/ACM Trans. Netw. vol. 19, no. 3, pp. 830–-840, June 2011.

\bibitem{Vermeulen2020NSDI}
K. Vermeulen, J. P. Rohrer, R. Beverly, O. Fourmaux, and T. Friedman, ``Diamond-Miner: Comprehensive discovery of the Internet{\textquoteright}s topology diamonds,'' in Proc. USENIX NSDI'20, pp. 479--493.

\bibitem{Almeida2020INFOCOM}
R. Almeida, I. Cunha, R. Teixeira, D. Veitch, and C. Diot, ``Classification of load balancing in the Internet,'' in Proc. IEEE INFOCOM'20, pp. 1987--1996.

\bibitem{Almeida2017PAM}
R. Almeida, O. Morais, E. Fazzion, D. Guedes, W. Meira Jr., and I. Cunha, ``A Characterization of Load Balancing on the IPv6 Internet,'' in Proc. PAM'17, pp. 242--254.


\bibitem{RIPE2015IPJ} 
RIPE NCC Staff, ``RIPE Atlas: A global Internet measurement network,'' The Internet Protocol Journal. vol. 18, no. 3 pp. 2–-26, 2015.

\bibitem{Giotsas2021TON}
V. Giotsas, G. Nomikos, V. Kotronis, P. Sermpezis, P. Gigis, L. Manassakis, C. Dietzel, S. Konstantaras, and X. Dimitropoulos, ``
O peer, where art thou? Uncovering remote peering interconnections at IXPs,'' IEEE/ACM Trans. Netw. vol. 29, no. 1 16 pages, Feb. 2021.



\bibitem{Javed2013SIGCOMM}
U. Javed, I. Cunha, D. R. Choffnes, E. Katz-Bassett, T. Anderson, and A. Krishnamurthy, ``PoiRoot: Investing the root cause of interdomain path changes'', in Proc. ACM SIGCOMM'13, pp. 183--194.

\bibitem{Comarela2013IMC} 
G. Comarela, G. G{\"u}rsun, and M. Crovella, ``Studying interdomain routing over long timescales,'' in Proc. ACM IMC'13, pp. 227–-234.

\bibitem{Paxson1997TON} 
V. Paxson,  ``End-to-end routing behavior in the Internet,'' IEEE/ACM Trans. Netw. vol. 5, no. 5, pp. 601-–615, Oct. 1997.

\bibitem{Fanou2015PAM} 
R. Fanou, P. Francois, and E. Aben,  ``On the diversity of interdomain routing in Africa,'' in PAM'15, J. Mirkovic and Y. Liu (Eds.). Springer International Publishing, pp. 41-–54.

\bibitem{Medem2012INFOCOM} 
A. Medem, C. Magnien, and F. Tarissan, ``Impact of power-law topology on IP-level routing dynamics: Simulation results,'' in Proc. IEEE INFOCOM'12, pp. 220–-225.

\bibitem{Rimondini2014PAM} 
M. Rimondini, C. Squarcella, and G. Di Battista, ``Towards an automated investigation of the impact of BGP routing changes on network delay variations,'' in Proc. PAM'14, pp. 193-–203.


\bibitem{Ahmed2015LCN} 
N. Ahmed, and K. Sarac, ``An experimental study on inter-domain routing dynamics using IP-level path traces,'' in Proc. IEEE ICN'15, pp. 510–-517.

\bibitem{Wassermann2017BigDAMA} 
S. Wassermann, P. Casas, T. Cuvelier, and B. Donnet, ``NETPerfTrace: Predicting Internet path dynamics and performance with machine learning,'' in Proc. ACM Big-DAMA'17, pp. 31-–36.

\bibitem{Cunha2014TON} 
I. Cunha, R. Teixeira, D. Veitch, and C. Diot, ``DTRACK: A system to predict and track Internet path changes,'' IEEE/ACM Trans. Netw. vol. 22, no. 4 pp. 1025–-1038, Aug. 2014.

\bibitem{Singh2015IEEECST} 
S. K. Singh, T. Das, and A. Jukan, ``A survey on Internet multipath routing and provisioning,'' IEEE Commun. Surv. Tutor. vol. 17, no. 4, pp. 2157–-2175, fourthquarter 2015.

\bibitem{Qadir2015IEEECST} 
J. Qadir, A. Ali, K. A. Yau, A. Sathiaseelan, and J. Crowcroft, ``Exploiting the power of multiplicity: A holistic survey of network-layer multipath,'' IEEE Commun. Surv. Tutor. vol. 17, no. 4, pp. 2176-–2213, fourthquarter 2015.

\bibitem{Augustin2007E2EMON}
B. Augustin, T. Friedman, and R. Teixeira, ``Multipath tracing with Paris traceroute,'' in Proc. IEEE E2EMON'07, pp. 1--8.

\bibitem{Veitch2009INFOCOM}
D. Veitch, B. Augustin, R. Teixeira, and T. Friedman, ``Failure control in multipath route trace,'' in Proc. IEEE INFOCOM'10, pp. 1395--1403.

\bibitem{Augustin2006IMC} 
B. Augustin, X. Cuvellier, B. Orgogozo, F. Viger, T. Friedman, M. Latapy, C. Magnien, and R. Teixeira, ``Avoiding traceroute anomalies with Paris traceroute,'' in Proc. ACM IMC'06, pp. 153–-158.

\bibitem{Valera2011MBGP}
F. Valera, I. van Beijnum, A. Garcia-Martinez, and M. Bagnulo, ``Multi-path BGP: Motivations and solutions,'' in Next-Generation Intenet Architectures and Protocols, B. Ramamurthy, G. N. Rouskas, and K. M. Sivalingam, Ed. Cambrige, UK: Cambridge Univ. Press, 2011.

\bibitem{Li2020TMA}
J. Li, V. Giotsas, and S. Zhou, ``Anatomy of BGP-Multipath deployment in a large ISP network,'' in Proc. TMA'20, arXiv: http://arxiv.org/abs/2012.07730.

\bibitem{Li2021GI}
J. Li, S. Zhou, and V. Giotsas, ``Performance analysis of BGP-Multipath,'' in Proc. IEEE GI'21, arXiv: http://arxiv.org/abs/2103.07683.

\bibitem{Fujinoki2008ICON}
H. Fujinoki,  ``Multi-Path BGP (MBGP): A solution for improving network bandwidth utilization and defense against link failures in inter-domain routing,'' in Proc. IEEE ICON'08, pp. 1–-6. 

\bibitem{Beijnum2009INFOCOM}
I. van Beijnum, J. Crowcroft, F. Valera, and M. Bagnulo. ``Loop-freeness in BGP-Multipath through propagating the longest path,'' in Proc. IEEE INFOCOM'09, 6 pages.

\bibitem{Huffaker2010PAM}
B. Huffaker, A. Dhamdhere, M. Fomenkov, and k claffy, ``Toward topology dualism: Improving the accuracy of AS annotations for routers,'' in PAM'10, A. Krishnamurthy and B. Plattner (Eds.). Springer International Publishing, pp. 101--110.

\bibitem{Marder2018IMC} 
A. Marder, M. Luckie, A. Dhamdhere, B. Huffaker, kc claffy, and J. M. Smith, ``Pushing the boundaries with bdrmapIT: Mapping router ownership at Internet scale,'' in Proc. ACM IMC'18, pp. 56-–69. 

\bibitem{Nur2018Comput.Netw.} 
A. Y. Nur, and M. E. Tozal, ``Cross-AS (X-AS) Internet topology mapping,'' Comput. Netw. vol. 132, pp. 53-–67, 2018.

\bibitem{Marder2016IMC} 
A. Marder, and J. M. Smith, ``MAP-IT: Multipass accurate passive inferences from traceroute,'' in Proc. ACM IMC'16, pp. 397-–411. 

\bibitem{RFC2992}
C. Hopps,  Analysis of an Equal-Cost Multi-Path Algorithm. RFC 2992, November 2000.

\bibitem{IETF-ECMP}
P. Lapukhov, Equal-cost Multipath Considerations for BGP, Internet Engineering Task Force. Network Working Group Internet Draft. https://tools.ietf.org/id/draft-lapukhov-bgp-ecmp-considerations-02.html. July 2019.


\bibitem{Mok2018PAM} 
R. K. P. Mok, V. Bajpai, A. Dhamdhere, and K. C. Claffy,  ``Revealing the load-balancing behavior of YouTube traffic on interdomain links,'' in PAM'18, R. Beverly, G. Smaragdakis, and A. Feldmann (Eds.), pp. 228–-240.

\bibitem{RFC7947}
E. Jasinska, N. Hilliard, R. Raszuk, and N. Bakker, ``Internet exchange BGP route server,'' RFC 7947, September 2016.

\bibitem{Pansiot2010PAM}
J.-J. Pansiot, P. Mérindol, B. Donnet, and O. Bonaventure, ``Extracting intra-domain topology from mrinfo probing,'' in PAM'10, A. Krishnamurthy and B. Plattner (Eds.). Springer International Publishing, pp. 81--90.

\bibitem{Giotsas2015CoNEXT} 
V. Giotsas, G. Smaragdakis, B. Huffaker, M. Luckie, and kc, claffy, ``Mapping peering interconnections to a facility,'' in Proc. ACM CoNEXT'15, Article No. 37.

\bibitem{mi2}
R, Motamedi, B. Yeganeh, B. Chandrasekaran, R. Rejaie, B. M. Maggs, and W. Willinger. ``On mapping the interconnections in today’s Internet,'' IEEE/ACM Trans. Netw., vol. 27, no. 5, pp. 2056--2070, 2019.

\bibitem{Luckie2016IMC} 
M. Luckie, A. Dhamdhere, B. Huffaker, D. Clark, and kc claffy,  ``Bdrmap: Inference of borders between IP networks,'' in Proc. ACM IMC'16, pp. 381–-396. 

\bibitem{Yeganeh2019IMC}
B. Yeganeh, R. Durairajan, R. Rejaie, and W. Willinger, ``How cloud traffic goes hiding: A study of Amazon's peering fabric,'' in Proc. ACM IMC'19, pp. 202--216.


\bibitem{RouteViews} 
University of Oregon Route Views Project,
\url{http://www.routeviews.org/}. (January 2020).

\bibitem{RIPERIS}
RIPE Routing Information Service. \url{http://www.ripe.net/ris}

\bibitem{Khan2013IMC}
A. Khan, T. T. Kwon, H.-C. Kim, and Y. Choi, ``AS-level topology collection through looking glass servers,'' in Proc. ACM IMC'13, pp. 235--241.

\bibitem{Chang2004Comput.Netw.}
H. Chang, R. Govindan, S. jamin, S. J. Shenker, and W. Willinger, ``Towards capturing representative AS-level Internet topologies,'' Comput. Netw., vol. 44, pp. 737--755, 2004.

\bibitem{Han2006TDS}
J. Han, D. Watson, and F. Jahanian, ``An experimental study of Internet path diversity,'' IEEE Trans. Dependable and Secure Computing, vol. 3, no. 4, pp. 273--288, 2006. 

\bibitem{Zhang2005CCR}
B. Zhang, R. Liu, D. Massey, and L. Zhang, ``Collecting the Internet AS-level topology,'' ACM SIGCOMM CCR, vol. 35, no. 1, pp. 53--62, 2005.

\bibitem{Holbert2015TNSM}
B. Holbert, S. Tati, S. Silvestri, T. F. La Prota, and A. Swami, ``Network topology inference with partial information'', IEEE TNSM, vol. 12, no. 3, pp. 406--419, 2015.

\bibitem{Giotsas2016PAM}
V. Giotsas, A. Dhamdhere, and {kc} claffy, ``Periscope: Unifying looking glass querying,'' in Proc. PAM'16, pp. 177--189.

\bibitem{BGPLGDatabase}
BGP Looking Glass Databases, http://www.bgplookingglass.com/. (January 2020)

\bibitem{PeeringDBAPI}
PeeringDB API Documentation, https://www.peeringdb.com/apidocs/. (January 2020)

\bibitem{PeeringDB}
PeeringDB, https://www.peeringdb.com/. (July 2021)

\bibitem{traceroute}
traceroute.org, http://traceroute.org/. (January 2020)

\bibitem{caida_asrank} 
CAIDA AS Rank, http://as-rank.caida.org/. (January 2020).

\bibitem{Luckie2013IMC}
M. Luckie, B. Huffaker, A. Dhamdhere, V. Giotsas, and kc claffy, ``AS relationships, customer cones, and validation,'' in Proc. ACM IMC'13, pp. 243--256.

\bibitem{hyper-giants}
T. B{\"o}ttger, C. Felix, and U. Steve, ``Looking for hypergiants in peeringDB,'' ACM SIGCOMM CCR, vol. 48, no. 3 (2018): 13-19.


\bibitem{HELG}
Looking Glass - Hurricane Electric (AS6939).
https://lg.he.net/


\bibitem{Ark} 
CAIDA: Archipelago (Ark) Measurement Infrastructure,
http://www.caida.org/projects/ark/.

\bibitem{Madhyastha2006OSDI} 
H. V. Madhyastha, T. Isdal, M. Piatek, and C. Dixon, ``iPlane: An information plane for distributed services,'' in  Proc. USENIX OSDI'06, pp. 367-–380.

\bibitem{RIPEstat}
RIPEstat Data API, 
https://stat.ripe.net/docs/data\_api\#whois

\bibitem{CISCO-LB}
CISCO, ``IP Switching Cisco Express Forwarding Configuration Guide, Cisco IOS XE Release 3S'',
https://www.cisco.com/c/en/us/td/docs/ios-xml/ios/ipswitch\_cef/configuration/xe-3s/isw-cef-xe-3s-book/isw-cef-load-balancing.html\#GUID-D8A86BB9-FCA8-48CA-881D-153F4383728D

\bibitem{Dhamdhere2018SIGCOMM}
A. Dhamdhere, D. D. Clark, A. Gamero-Garrido, M. Luckie, R. K. P. Mok, G. Akiwate, K. Gogia, V. Bajpai, A. C. Snoeren, and kc claffy, ``Inferring persistent interdomain congestion'', in Proc. ACM SIGCOMM'18, 15 pages. 

\bibitem{Koch2021HotNets}
T. Koch, W. Jiang, T. Luo, P. Gigis, Y. Zhang, K. Vermeulen, E. Aben, M. Calder, E. Katz-Bassett, L. Manassakis, G. Smaragdakis, and N. Vallina-Rodriguez, ``Towards a traffic map of the Internet'', in Proc. ACM HotNets'21, pp. 23-30.

\bibitem{GitHub}
GitHub-jieliucl/BGP-M.
https://github.com/jieliucl/BGP-M


















































\end{thebibliography}
\end{document}